# PRAXIS: Case-distilled and code-verified AI agents for biological research

Zhenyu Ma[1], Yuyang Song[1], Chunyi Yang[2], Jingyi Zhu[2], Limei Xu[2], Min Xiao[1], Xukai Jiang[1*]

[1]National Glycoengineering Research Center, Shandong University, Qingdao 266237, China.

[2]State Key Laboratory of Microbial Technology, Shandong University, Qingdao 266237, China

***Correspondence**

Xukai Jiang, Email: xukai.jiang@sdu.edu.cn

**Keywords:** Scientific agent; biocomputing; case-based learning; long-term memory; negative-case knowledge; code validation; adaptive retrieval; auditable workflow; PRAXIS

## Abstract

Large language models are moving scientific research from text assistance toward agentic workflows, yet biological research requires strong object validation, methodological suitability, reproducibility, and auditability. Prompt engineering, general RAG, or tool use alone cannot reliably produce domain-specific scientific judgment. Here, we present PRAXIS, a verifiable biological research agent framework driven by literature learning and case distillation. PRAXIS converts research experience, failure boundaries, domain rules, and executable procedures into structured long-term memory. By coordinating successful cases, negative cases, rules, and skills, PRAXIS supports problem definition, object validation, method selection, workflow execution, result interpretation, and review feedback across diverse biocomputational tasks. We instantiated PRAXIS as an agent suite for biomedical computing and evaluated it through object validation, case retrieval, memory ablation, public benchmarks, and cross-agent workflows. The results show that case-based learning improves method selection, error suppression, and workflow organization in complex biological research tasks. Rather than replacing scientists, PRAXIS provides a general pathway for transforming research experience into executable, auditable, and transferable agent capabilities.

## Introduction

Large language models (LLMs) are rapidly reshaping the organization of scientific research[1]. Through natural-language understanding, code generation[2], tool use, and long-context reasoning[3], LLMs can now support literature search, experimental design, data analysis, molecular design[4], and result interpretation. When connected to external tools, file systems, databases, and execution environments, they begin to function not merely as text generators but as scientific agents capable of organizing complex research workflows[5]. In biomedicine, this trend is evident across chemical synthesis, drug discovery[6], omics analysis[7], protein modeling[8], molecular dynamics[9], and gene-editing design[10].

However, biological research imposes reliability requirements far beyond ordinary text tasks. A useful biocomputational agent must not only produce plausible answers, but also correctly handle database identifiers, software versions, species and tissue contexts, experimental constraints, statistical assumptions, tool boundaries, and result provenance. Incorrect DOI, ChEMBL, PDB, GEO, or gene identifiers can directly point to the wrong paper, molecule, structure, dataset, or biological object[16]. Likewise, using cell-level Wilcoxon tests for single-cell differential expression[11], drawing conclusions from a single docking score[12], or ignoring force fields[13], membrane environments[14], and sampling timescales in molecular dynamics[15] can turn fluent responses into irreproducible or misleading scientific claims[17]. The central challenge is therefore not to make LLMs generate more content, but to enable verifiable, executable, and auditable domain-specific judgment.

Existing strategies only partially address this challenge. Prompt engineering can instruct models not to fabricate references or to call tools, but it cannot ensure that they know which objects require validation, which methods are inappropriate, or which routes should be rejected. Retrieval-augmented generation (RAG)[18] provides external text, yet similarity in biological research often depends on species, tissue, data type, experimental design, toolchain, structural state, chemical scaffold, or analytical purpose rather than surface semantic similarity. Tool calling improves executability[19], but tools themselves do not tell the model when to call them, how to combine them, when to stop, or how to interpret failure. Connecting an LLM to tools does not automatically make it a trained domain researcher.

We argue that reliable biological research agents require a paradigm closer to real scientific training. Researchers become competent not by reading a single prompt, but by repeatedly studying cases, extracting rules, executing workflows, encountering failures, undergoing review[20], and revising their judgment. In biomedicine, much critical knowledge is embedded in case experience rather than explicit rules: which catalytic residues should not be mutated, which scaffolds should be excluded because of intellectual-property or ADMET risks, which single-cell annotations require dual-source validation, and which in silico predictions can only support triage rather than replace wet-lab validation. Such negative knowledge and boundary conditions are rarely the main results of papers, but they are essential for reliable scientific agents.

Here, we present PRAXIS, a verifiable biological research agent framework driven by literature learning and case distillation. PRAXIS converts biological research experience into agent capabilities that can be accumulated, retrieved, reused, and audited. Unlike systems that rely only on prompts or general RAG, PRAXIS organizes experience as structured long-term memory. Literature-learning cases provide domain background, standard procedures, and tool boundaries; practice cases record task design, execution paths, and result interpretation; negative cases preserve failed routes and non-transferable conditions; rules distill stable cross-case constraints; and skills convert repeatedly validated procedures into executable workflow templates. In this way, PRAXIS learns not only what should be done, but also which paths should not be transferred.

At runtime, PRAXIS retrieves relevant cases, negative cases, rules, and skills according to the user's biological research question, supporting task definition, protocol design, execution planning[21], result reporting, and review feedback. Successful cases provide transferable routes, skills provide executable procedures, rules define domain boundaries, and negative cases suppress known failure modes. Each stage is constrained by schemas, reviewer gates, and code-validation mechanisms. Key numerical values, database objects, and intermediate conclusions must be supported by tool execution, database queries, or result files rather than model self-assertion. To cover the diversity of computational biology tasks, we instantiated PRAXIS as a suite of agents spanning 13 directions: molecular dynamics, enzyme engineering, omics, microbiome analysis, cell analysis, protein modeling, molecular docking, cryo-EM, drug discovery, CRISPR, immunoinformatics, pharmacology, and synthetic biology. All agents share the same case-learning grammar while retaining domain-specific validators, retrieval channels, failure modes, and tool backends. The current PRAXIS canonical brain contains 278 cases, 110 negative cases, 156 rules, and 150 executable skill workflows.

We evaluated whether PRAXIS can convert case learning into stable agent capability through object validation, case retrieval, memory ablation, public benchmarks, and cross-agent workflows. Code-enforced identifier QA tested whether biological identifiers generated by the model could be resolved in real databases and whether incorrect objects could be repaired or blocked. A 110-case leave-one-out retrieval experiment evaluated whether long-term case memory could be correctly recalled for new tasks. Memory ablation further separated the contributions of rules, skills, cases, and negative cases to reducing unsafe recommendations.

To move beyond internal memory tests, we evaluated rule-guided method selection on public Drug, CRISPR, and Omics benchmarks. Finally, schema-validated envelopes and four curated cross-agent workflow traces demonstrated how PRAXIS organizes multiple domain agents into routable, verifiable, and auditable workflows. These traces are intended to validate system-level coordination rather than serve as evidence of autonomous biological discovery.

Overall, PRAXIS does not aim to replace scientists with an autonomous discovery system. Instead, it provides a general approach for transforming real scientific experience into executable, auditable, and transferable agent capabilities. The following Results sections present the PRAXIS architecture, identifier QA and adaptive retrieval, memory ablation, public benchmark validation, and schema-envelope-supported cross-agent workflows.

## Results

### PRAXIS is an executable biological research agent suite driven by literature learning and case distillation

PRAXIS was designed as an executable agent suite for biological research tasks. Because complex bioinformatics studies often involve diverse data objects, software environments, and multi-step analytical decisions, we instantiated PRAXIS across 13 biocomputational directions: molecular dynamics, enzyme engineering, omics, microbiome analysis, cell analysis, protein modeling, molecular docking, cryo-EM, drug discovery, CRISPR, immunoinformatics, pharmacology, and synthetic biology (Figure 1a). These agents differ in input objects, toolchains, and domain-specific failure modes. For example, PRAXIS-MD must handle force fields, membrane systems, sampling strategies, and trajectory analysis; PRAXIS-Omics must process count matrices, h5ad objects, batch effects, and differential analysis designs; PRAXIS-Drug must validate chemical identifiers and select appropriate screening or validation strategies; and PRAXIS-CRISPR must evaluate guide RNAs, off-target candidates, and editing windows. Despite these differences, all PRAXIS agents share the same learning grammar: tasks are supported by prior cases, cases inform rules, rules constrain executable skills, skills generate auditable evidence through execution, and validated results are written back to long-term memory.

The PRAXIS long-term brain is not a simple accumulation of model-generated free text. It is built from two complementary sources of experience. The first is literature-learning experience, which provides domain background, standard procedures, tool boundaries, database objects, and key terminology through systematic curation of papers, database documentation, software manuals, and field guidelines. The second is practice-case experience, distilled from real research-style agent interactions in which domain scientists manually guided, reviewed, and corrected task scoping, method selection, tool outputs, failure causes, and result interpretation.

The current PRAXIS canonical brain contains 278 cases, 110 negative cases, 156 rules, and 150 executable skill workflows across 13 biological computing agents (Figure 1b). Cases include both literature-learning cases that define domain knowledge and methodological boundaries, and practice cases generated through scientist-guided task execution and review. Negative cases record failed routes, invalid identifiers, inappropriate method transfer, and conditions requiring human inspection. Rules distill stable patterns repeatedly observed across cases, whereas skills convert reusable procedural knowledge into executable workflows.

Conceptually, PRAXIS represents biological research experience as interconnected computable objects rather than unstructured text (Figure 1c). User tasks first retrieve relevant cases and negative cases. These cases are not ordinary few-shot examples or one-off model outputs, but reviewed research precedents that preserve scientific context, input objects, tool choices, parameter settings, quality-control criteria, exclusions, and failure routes. Recurrent patterns across multiple cases can be promoted into rules, but observations from a single case are not treated as universal principles. Rules then constrain skills by specifying when a method is applicable, when it should be avoided, and which checks are required before outputs can support conclusions. Through these object relationships, PRAXIS organizes experience into reusable links among task, case, negative case, rule, skill, execution, validation, report, and memory.

This object-level representation is critical for biological research agents because similarity in biological tasks is rarely reducible to textual similarity. An omics task may resemble a prior case because of tissue source, perturbation design, sample structure, or annotation strategy. A drug discovery task may require experience retrieval based on target class, chemical scaffold, assay type, or ADMET risk. A molecular dynamics task may depend on structural state, membrane composition, force-field choice, sampling strategy, and trajectory preprocessing. By storing cases as structured research precedents, PRAXIS enables retrieval around scientific context rather than surface wording. By promoting rules only after repeated cross-case support, it reduces the risk of overgeneralizing single observations. By translating rules into skills, it forces procedural knowledge to become executable rather than remaining as textual advice.

At runtime, these objects form a repairable scientific workflow (Figure 1d). Each task begins with a user query. During scoping, the system identifies the domain, task type, input constraints, and required evidence. Case retrieval then searches long-term memory[22] for relevant precedents, negative cases, and rules. During planning, the retrieved knowledge is used to generate candidate protocols and select appropriate skills or toolchains. During execution, concrete tools and skills are run, rather than allowing the LLM to directly fabricate values or conclusions. The resulting files, parameters, logs, and intermediate outputs are then passed to QA and validation, where identifiers, schemas, tool outputs, parameter ranges, and domain-specific sanity criteria are checked. Only validated results enter the reporting stage and support the final answer.

PRAXIS also includes an explicit reviewer gate that combines schema checks, domain rules, negative-case memory, and repair hints. If a plan lacks required fields, parameter choices conflict with retrieved cases, tool execution fails, or validation detects unreliable outputs, the system does not silently proceed to reporting. Instead, it triggers replanning, re-execution, or low-confidence labeling. In this way, errors are exposed, repaired, or bounded rather than hidden by fluent language. This mechanism mimics the review-and-revision process of real scientific training[23] and subjects the agent's conclusions to recorded quality control.

The final step is memory update. When execution and validation produce reliable results, the experience can be written back into long-term memory as a new practice case, an update to an existing case, supporting evidence for a rule, or an improvement to an executable skill. Conversely, failed tasks are also retained as negative experience, including invalid identifiers, unsuitable tool routes, non-transferable parameters, failed assumptions, or boundary conditions requiring human review. Thus, PRAXIS does not learn by modifying the parameters of the underlying LLM. Instead, it accumulates cases, rules, skills, and failure boundaries in persistent research memory, allowing subsequent tasks to reuse verified experience.

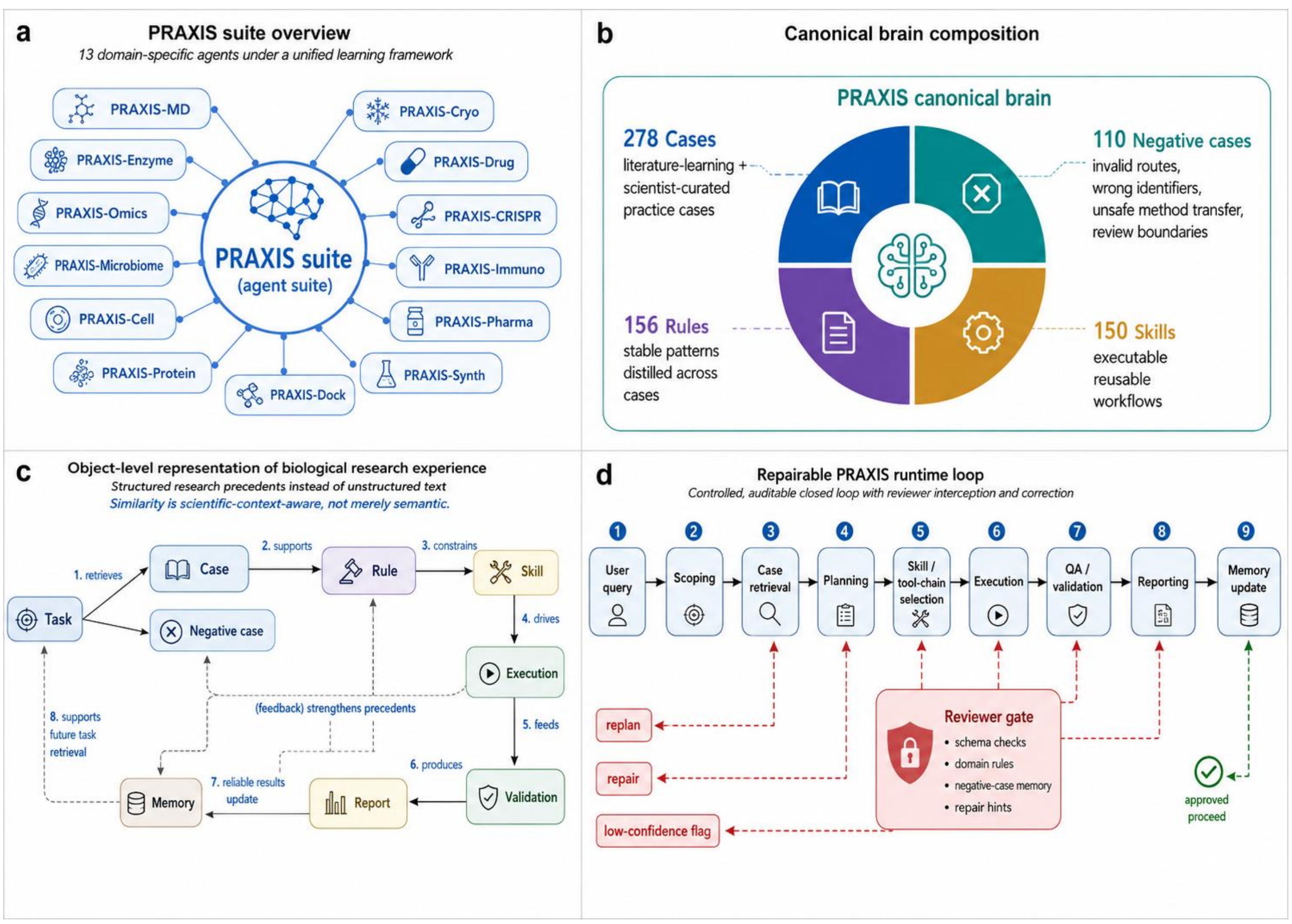

**Figure 1. PRAXIS is an executable biological research agent suite driven by literature learning and case distillation.**

**(a)** Overview of the PRAXIS suite, which integrates 13 domain-specific agents, including PRAXIS-MD, PRAXIS-Enzyme, PRAXIS-Omics, PRAXIS-Microbiome, PRAXIS-Cell, PRAXIS-Protein, PRAXIS-Dock, PRAXIS-Cryo, PRAXIS-Drug, PRAXIS-CRISPR, PRAXIS-Immuno, PRAXIS-Pharma, and PRAXIS-Synth. **(b)** Composition of the PRAXIS canonical brain, containing 278 cases, 110 negative cases, 156 rules, and 150 executable skill workflows. Cases provide literature-derived and scientist-refined experience; negative cases record failed routes and non-transferable conditions; rules capture stable cross-case constraints; and skills encode executable procedures. **(c)** Object-level representation of research experience. PRAXIS links tasks, cases, negative cases, rules, skills, execution, validation, reports, and memory into a reusable structure for biological research workflows. **(d)** Repairable PRAXIS execution loop. Each task proceeds through scoping, case retrieval, planning, skill/tool execution, QA/validation, reporting, and memory update. A reviewer gate detects missing fields, tool failures, unreliable outputs, or inappropriate method transfer, triggering replanning, repair, or low-confidence labeling.

**Code validation and adaptive retrieval turn PRAXIS long-term memory into executable experience**

After establishing the PRAXIS architecture, we examined whether its long-term brain could be safely and accurately reused in new biological research tasks. For a case-learning scientific agent, experience reuse requires more than retrieving related examples. Biological objects entering the workflow must first be real, resolvable, and semantically consistent with the task; then, the system must retrieve genuinely transferable prior experience rather than being misled by surface wording, weak tags, or irrelevant toolchains. We therefore evaluated PRAXIS long-term memory at two levels: code-enforced identifier QA to block incorrect objects from downstream execution, and adaptive case retrieval to recall appropriate historical experience (Figure 2a).

We first assessed the risk of biological identifier errors. In biocomputing workflows, identifiers such as ChEMBL[24] IDs, SMILES, DOIs, UniProt[26] accessions, PDB IDs, EMDB[28] accessions, HGNC[29] symbols, Ensembl[30] IDs, and GEO accessions are not ordinary text: they directly determine database retrieval, structure download, molecule preparation, sequence mapping, and downstream analysis. To measure this risk, we asked Claude Opus 4.7 to generate 80 structured biology case YAML files across drug, cryo-EM, protein, CRISPR, and omics domains without external APIs or tools. PRAXIS then extracted key identifiers and independently validated them through ChEMBL, PubChem[25], CrossRef, UniProt, RCSB-PDB[27], EMDB, HGNC, Ensembl, NCBI, or code-based interfaces. Each object was classified as executable, repairable, or unsafe.

Identifier hallucination showed strong domain specificity. Drug cases had the highest error rate: 13 of 30 compound records contained identifier errors, corresponding to a hallucination rate of 43.3%. Cryo-EM showed an intermediate error rate of 30.0% with 3 errors in 10 cases. In contrast, protein, CRISPR, and omics each showed 0.0% identifier errors in 10 cases (Figure 2b). These results suggest that LLM memory is uneven across biological identifier types. Chemical IDs, tautomers, stereochemical forms, and structured chemical representations are particularly vulnerable to weak semantic binding errors, whereas gene, protein, and omics accessions were more stable in this sample.

In the drug domain, the 13 errors mainly reflected unstable binding between chemical entities and database objects. They included six wrong ChEMBL ID or tautomer cases[31], four stereochemical descriptor differences[32], two invalid DOIs, and one nonexistent ChEMBL ID. These errors indicate that the model may recall drug names, related literature, or approximate structures, while failing to distinguish primary ChEMBL records, tautomeric states, stereochemical forms, and resolvable accessions. Thus, chemical-object errors often appear not as fully unrelated hallucinations, but as locally related yet execution-invalid object binding failures. Code-enforced QA converted these errors into actionable program events. For the 13 chemistry-side errors, PRAXIS achieved 13/13 automatic repair through ChEMBL search-by-name fallback, yielding a 100% auto-correction rate. Twelve errors were corrected by SMILES or tautomer replacement, and one was corrected by ChEMBL ID replacement (Figure 2c). Spot-checking showed that 12 of 13 repairs were semantically acceptable, none had a connectivity-level mismatch, and one case was routed to human review because of an ambiguous tautomer boundary.

System-level baseline analysis further showed the protective role of identifier QA. Among 30 drug identifier cases, removing identifier QA would allow 33.3% of cases to enter downstream workflows with wrong-object or stereo-risk status. With code-verified QA, the pass-or-repair rate increased to 96.7%, and unreviewed semantic mismatch decreased to 0.0% (Figure 2d). Thus, identifier QA functions not only as an error-checking module, but also as a memory protection mechanism: it prevents incorrect objects from entering execution chains and being written back, retrieved, or reused in future tasks. Beyond object validation, PRAXIS must also retrieve the right prior cases. In biological research, case similarity is rarely determined by surface text alone. Two tasks may share similar wording but differ in species, tissue source, data structure, chemical scaffold, structural state, tool applicability, or failure boundary; conversely, two differently worded tasks may share transferable methodological constraints or risk patterns. PRAXIS therefore treats case retrieval as a core condition for case learning, rather than as ordinary RAG-style text recall.

We constructed a 110-case leave-one-out retrieval benchmark. In each run, one case was used as the held-out query, represented only by its frontmatter and an approximately 200-token summary, while the remaining 109 cases formed the retrieval library. Gold similarity was defined as other cases from the same agent. We compared BM25[33], Tags Jaccard, Tools Jaccard, naive RRF[34], weighted RRF, and PRAXIS adaptive retrieval. PRAXIS adaptive retrieval achieved the best recall@10 of 0.683, with a 95% CI of 0.64–0.72, outperforming BM25 alone at 0.646, 95% CI 0.61–0.68, paired Wilcoxon $p = 0.009$. More importantly, adaptive retrieval substantially outperformed naive multi-channel fusion: RRF naive reached only 0.360, 95% CI 0.31–0.42, $p = 5.4 \times 10^{-16}$, and RRF weighted reached 0.326, 95% CI 0.27–0.38, $p = 1.0 \times 10^{-16}$ (Figure 2e, S1).

This result reveals that multi-channel fusion is not automatically superior to a strong single channel. In a cross-domain biological case library, signal strength differs across retrieval channels. Equal or near-equal fusion of text, tags, and tools can dilute the strongest signal and push relevant cases lower in the ranking. The key retrieval question in PRAXIS is therefore not whether more channels are used, but which similarity signal should be trusted for the current task. Per-agent analysis supported this interpretation. In cryo-EM, BM25 recall@10 reached 0.833, suggesting that domain-specific terminology was distinctive enough for text retrieval to separate relevant cases. In omics, Tools Jaccard reached a recall@10 of 1.000, because the 10 omics cases shared highly consistent tool ecosystems such as Scanpy[35], Seurat[36], and DESeq2[37]. Drug retrieval was more balanced, with BM25, Tags Jaccard, and Tools Jaccard all close to 0.5, indicating that chemical relevance depends jointly on text, tags, and tools. Synth showed weaker performance across methods, suggesting greater internal heterogeneity and the need for finer-grained features such as pathway, host, construct, and objective (Figure 2f, S1).

Accordingly, PRAXIS uses domain-aware adaptive channel selection rather than fixed RRF fusion. This reframes retrieval as task-specific method selection: the system must decide which prior cases are scientifically comparable, rather than mechanically combining similarity scores. It also explains why the PRAXIS long-term brain cannot be reduced to a single vector database. Different agents store reliable experience in different object structures; some tasks depend on terminology, others on toolchains, and others on task goals, failure modes, or execution boundaries.

Case-library scaling further supported the idea that memory accumulation progressively improves experience recall. As the available case-library fraction increased from 0 to 0.25, 0.50, 0.75, and 1.00, recall@10 increased monotonically from 0.0000 to 0.8000, 0.9450, 0.9820, and 1.0000, respectively (Figure 2g, S2). The largest gain occurred early: using only 25% of the case library already achieved a recall@10 of 0.8000. Later expansion produced diminishing returns, suggesting that once the case base reaches a sufficient scale, further improvement depends more on retrieval routing, deduplication, negative-case separation, and case-quality control than on simply adding more cases[38].

Together, these analyses show that PRAXIS makes long-term memory executable through two complementary mechanisms. Code-verified identifier QA ensures that database objects entering the workflow are real, resolvable, and semantically consistent, while converting chemical-object errors into repair or human-review events. Adaptive retrieval selects the most appropriate retrieval channel according to task domain and case structure, preventing weak signals from diluting stronger ones and ensuring that transferable cases enter planning and validation. Thus, PRAXIS case learning is not merely the storage of more text, but the transformation of long-term memory into executable, repairable, and auditable research experience through object validation, task-aware retrieval, and cumulative case learning.

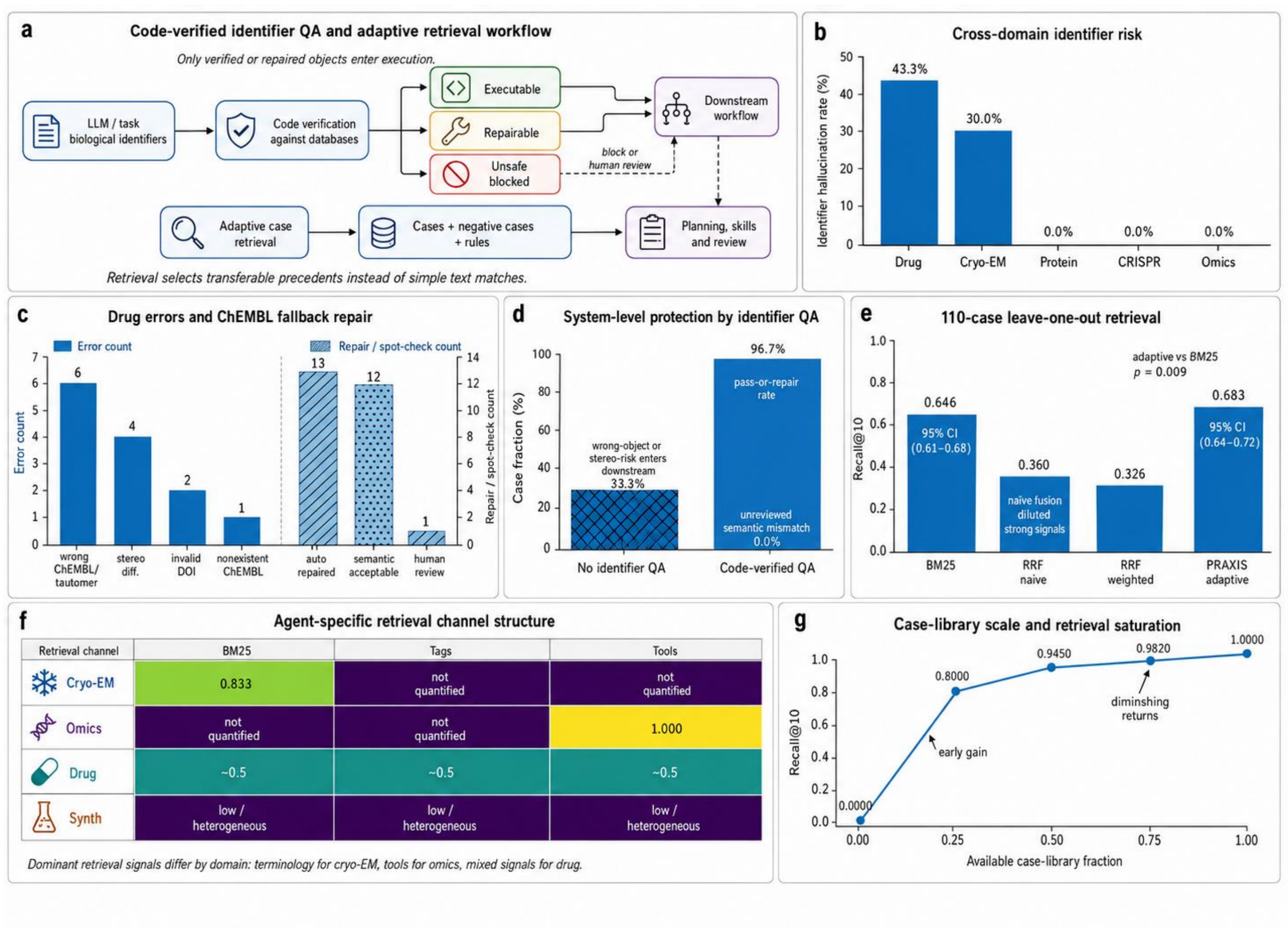

**Figure 2. Code validation and adaptive retrieval make PRAXIS long-term memory executable.** **(a)** Workflow of identifier QA and adaptive retrieval. PRAXIS validates biological identifiers as executable, repairable, or unsafe; only validated or repaired objects enter downstream workflows. Adaptive retrieval then recalls relevant cases, negative cases, and rules for planning, execution, and review. **(b)** Identifier hallucination rates across domains. Drug cases showed the highest error rate, with 13/30 compound records incorrect (43.3%). Cryo-EM showed 3/10 errors (30.0%), whereas protein, CRISPR, and omics showed no detected identifier errors. **(c)** Drug-domain error types and repair. The 13 errors included wrong ChEMBL ID or tautomer assignments (n = 6), stereochemical differences (n = 4), invalid DOIs (n = 2), and nonexistent ChEMBL IDs (n = 1). ChEMBL search-by-name fallback repaired all 13 errors; 12/13 repairs were semantically acceptable, with no connectivity-level mismatch and one case routed to human review. **(d)** System-level effect of identifier QA. Without identifier QA, 33.3% of drug cases would enter downstream workflows with wrong-object or stereo-risk status. With code-verified QA, the pass-or-repair rate reached 96.7%, and unreviewed semantic mismatch fell to 0.0%. **(e)** Leave-one-out retrieval benchmark. In 110 cases, PRAXIS adaptive retrieval achieved the best recall@10 of 0.683, outperforming BM25 (0.646; paired Wilcoxon $p = 0.009$), RRF naive (0.360), and RRF weighted (0.326), indicating that adaptive retrieval avoids dilution from weak channels. **(f)** Agent-specific retrieval signals. Cryo-EM relied mainly on BM25 (recall@10

= 0.833), omics on Tools Jaccard (recall@10 = 1.000), drug on mixed text, tag, and tool signals, while synth showed weaker performance due to higher case heterogeneity. **(g)** Case-library scaling. As the available case fraction increased from 0 to 0.25, 0.50, 0.75, and 1.00, recall@10 increased from 0.0000 to 0.8000, 0.9450, 0.9820, and 1.0000, showing that accumulated cases improve transferable-experience recall with diminishing later gains.

**Rules, skills, and negative-case knowledge jointly reduce unsafe scientific transfer**

After showing that PRAXIS can block incorrect objects through code validation and retrieve relevant cases through adaptive retrieval, we further tested whether its long-term brain could reduce a more subtle risk: unsafe transfer between superficially similar scientific tasks. Unlike identifier hallucination, this error does not usually appear as an unresolvable database object. Instead, the agent mechanically transfers methods, parameters, or interpretation boundaries from prior cases to new tasks where they are inappropriate. Such errors are especially dangerous in case-learning systems, because once a wrong route enters long-term memory, it may be retrieved, reused, and amplified in future tasks. We therefore evaluated the distinct contributions of rules, skills, cases, and negative cases to unsafe recommendation control (Figure 3a).

We designed a case-memory ablation experiment across 11 domain agents: cell, CRISPR, cryo, dock, drug, immuno, microbiome, omics, pharma, protein, and synth. Each agent included five held-out tasks independent of the long-term case library, yielding 55 tasks. Each task was run under six brain conditions: no_brain, cases_only, skills_only, rules_only, cases+NEG, and full_PRAXIS, producing 330 fixed outputs. The primary metric was deterministic unsafe-recommendation rate, scored by predefined string, keyword, and fuzzy-matching rules rather than an LLM judge, allowing us to evaluate scientific risk without conflating reliability with linguistic fluency (Figure S3). Unsafe_rate decreased as more PRAXIS memory components were introduced. The rate was 0.5818 under no_brain, 0.5455 under cases_only, 0.4909 under skills_only, 0.4364 under rules_only, 0.3636 under cases+NEG, and 0.2182 under full_PRAXIS.

Compared with no_brain, full_PRAXIS reduced unsafe_rate by 0.3636, corresponding to an approximately 62.5% relative reduction, with non-overlapping 95% confidence intervals. A paired discordant-pair sign test supported this effect: among 24 discordant tasks, full_PRAXIS outperformed no_brain in 22 and lost in 2, yielding a two-sided binomial $p \approx 3.6 \times 10^{-5}$. Paired comparisons showed that different memory components contributed distinct gains. Cases+NEG significantly outperformed no_brain, winning 16 of 20 discordant pairs, $p = 0.0118$, indicating that case memory with negative-case knowledge already reduces unsafe recommendations. Full_PRAXIS also outperformed skills_only, winning 18 of 21 discordant pairs, $p = 0.00149$, showing that a skill library alone cannot substitute for the full PRAXIS brain. Full_PRAXIS further improved final reviewer score relative to no_brain, winning 36 of 50 discordant pairs, p = 0.00260 (Figure 3b). These results indicate that PRAXIS reliability arises from the combined effect of rules, skills, cases, and negative cases.

The independent contribution of negative cases was assessed by comparing cases_only with cases+NEG. These two settings were identical except for the negative-knowledge blocks describing exclusions and failure modes. Unsafe_rate decreased from 0.5455 under cases_only to 0.3636 under cases+NEG, an absolute reduction of 0.1818 and a relative reduction of approximately 33.3% (Figure 3c). In paired testing, cases+NEG won 17 of 24 discordant tasks, $p \approx 0.064$. Although this did not cross the 0.05 threshold, the consistent direction, sizable effect, and agreement with the six-condition unsafe_rate gradient support a substantive role for negative-case knowledge in fine-grained failure avoidance.

This experiment also revealed the limitation of broad quality metrics. Plan_completeness, method_appropriateness, and final_reviewer_score were close to ceiling across conditions, and identifier_validation_rate was 1.000 in all settings. This suggests that frontier models such as Claude Opus 4.7[39] already perform well on coarse plan completeness and explicit identifier validation. In contrast, deterministic unsafe_rate captured finer failure modes, including ignored exclusion criteria, inappropriate method transfer, missing boundary statements, and superficially plausible but scientifically unsafe recommendations (Figure 3d, S3). Thus, the measurable value of case memory lies less in making answers appear more complete and more in reducing fine-grained scientific risk.

To test whether this effect depended on a single model, we repeated the six-condition experiment on cryo, drug, and omics subsets using Claude Sonnet and Claude Haiku, and compared them with Claude Opus. All three models showed the same overall pattern: no_brain had the highest unsafe_rate, whereas full_PRAXIS had the lowest or tied-lowest unsafe_rate (Figure 3e, S4). The reduction from cases_only to cases+NEG was also reproduced in Sonnet and Haiku, decreasing from 0.6000 to 0.3333 and from 0.6000 to 0.4000, respectively.

Worked examples showed how negative-case knowledge changed concrete planning decisions. In an EGFR screening task, no_brain tended to rely on a single AutoDock Vina redocking[40] route and ignored the structural distinction between EGFR-WT and the T790M resistance context[41]. Full_PRAXIS, after retrieving relevant failure modes, explicitly rejected the “Vina-only” shortcut, required Vina plus GNINA[42] dual-engine validation, and stated that EGFR-WT results could not substitute for the T790M mutant setting. Similar patterns appeared in other agents: full_PRAXIS added GPCR stabilizer and multibody refinement in cryo tasks, switched from small-molecule docking to AF-Multimer[43,44] for peptides longer than 20 amino acids, required at least five seeds and multi-model cross-validation in protein modeling, and enforced deep-learning segmentation with arcsinh transformation in cell analysis (Figure 3f).

Together, these results show that PRAXIS improves reliability not through a single prompt, skill library, or case memory alone, but through a risk-control layer composed of rules, skills, cases, and negative cases. Rules provide stable cross-case constraints, skills provide executable procedures, cases provide transferable research precedents, and negative cases preserve failed routes and non-transferable boundaries. Across 330 ablation outputs, the full PRAXIS brain significantly reduced unsafe recommendations, while negative cases provided additional observable risk suppression. Thus, the value of PRAXIS case learning is not merely improved response completeness, but measurable reduction of fine-grained scientific risk.

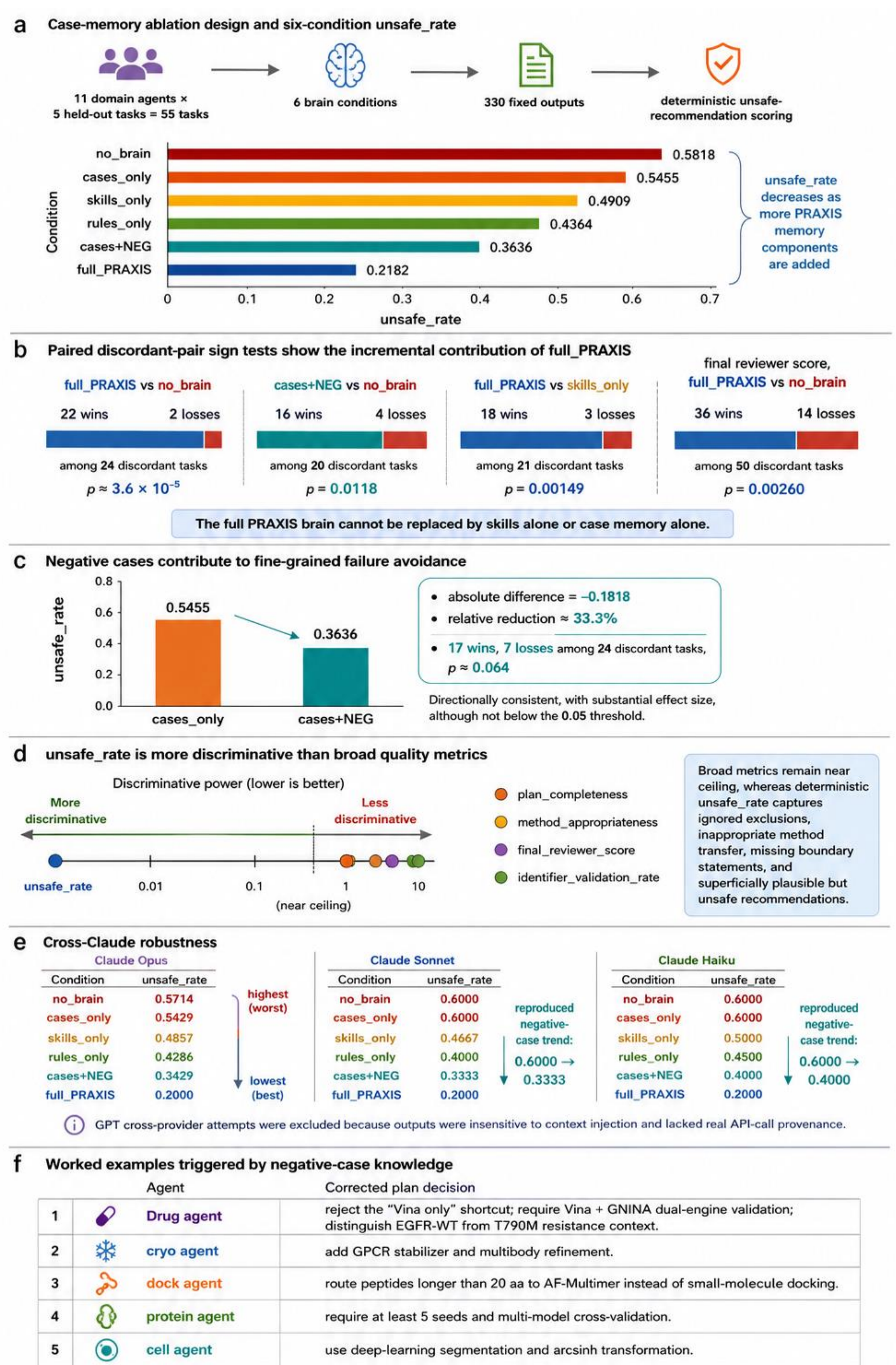

**Figure 3. Rules, skills, and negative cases reduce unsafe scientific transfer.**

**(a)** Case-memory ablation across 11 agents and 55 held-out tasks. Each task was tested under six brain conditions, producing 330 outputs. Unsafe_rate decreased from no_brain = 0.5818 to full_PRAXIS = 0.2182. **(b)** Discordant-pair sign tests showed significant gains from full_PRAXIS. Full_PRAXIS outperformed no_brain in 22/24 discordant tasks ($p \approx 3.6 \times 10^{-5}$), cases+NEG outperformed no_brain in 16/20 tasks ($p = 0.0118$), and full_PRAXIS outperformed skills_only in 18/21 tasks ($p = 0.00149$). **(c)** Negative cases reduced unsafe_rate from 0.5455 in cases_only to 0.3636 in cases+NEG, a relative reduction of about 33.3%. Although the paired test was marginal (17/24 wins, $p \approx 0.064$), the effect direction supported incremental risk suppression. **(d)** Unsafe_rate was more discriminative than broad quality metrics, which were near ceiling across conditions. It captured fine-grained risks such as inappropriate method transfer, missing exclusions, and insufficient boundary statements. **(e)** The trend was robust across Claude Opus, Sonnet, and Haiku in cryo, drug, and omics subsets: no_brain showed the highest unsafe_rate, while full_PRAXIS was lowest or tied-lowest. **(f)** Worked examples showed that negative cases changed concrete plans, including rejecting Vina-only EGFR screening, adding GPCR stabilizer and multibody refinement, routing long peptides to AF-Multimer, requiring multi-seed protein modeling, and enforcing deep-learning segmentation with arcsinh transformation.

**Public benchmarks support rule-guided adaptive method selection**

The internal ablation results showed that PRAXIS reliability does not arise from simply adding more memory or more tools, but from the coordinated effects of object validation, adaptive retrieval, rule constraints, and negative-case suppression. To test whether this principle generalizes to public tasks, we evaluated PRAXIS-Drug, PRAXIS-CRISPR, and PRAXIS-Omics on three representative biocomputational benchmarks: ligand-based virtual screening, CRISPR off-target prediction, and single-cell cell-type annotation. In these tasks, PRAXIS was implemented as a rule-guided adaptive method selector with online calibration, deciding when to use a strong single method, when to apply cross-validation, and when to restrict computational outputs to triage rather than final decisions.

In PRAXIS-Drug, we evaluated ligand-based virtual screening on six DUD-E targets[45], including EGFR, CDK2, ADRB2, ESR1, HIVPR, and ACE. For each target, 30 active molecules were sampled as leave-one-out queries, with the first five used for online calibration and the remaining 25 for testing. After calibration, PRAXIS-Drug selected Morgan ECFP4[46] for all six targets, fully matching the literature-prior rule. Across 150 test queries, PRAXIS-adaptive achieved a mean EF1% of 32.67 and a mean AUC of 0.819, reaching approximately 97.6% of the oracle upper bound of 33.48. In contrast, naive three-fingerprint consensus achieved a mean EF1% of only 23.61. Paired Wilcoxon testing showed that PRAXIS-adaptive significantly outperformed naive consensus, with a mean EF1% difference of +9.05 and $p = 2.4 \times 10^{-17}$ (Figure 4a). These results indicate that naive multi-method averaging can dilute the strongest molecular representation, whereas PRAXIS places method fusion itself under rule-guided and calibration-based review.

In PRAXIS-CRISPR, we used the Tsai 2015 GUIDE-seq[47] off-target benchmark to evaluate in silico off-target prediction and its safety boundary. The benchmark included four sgRNAs from VEGFA-T1, VEGFA-T3, EMX1, and FANCF. For each sgRNA, 32,551 candidate sites with ⩽3 mismatches were enumerated and scored using Hsu/MIT[48], CFD[49], and PRAXIS-adaptive scoring. PRAXIS-adaptive matched the best single method, Hsu/MIT, with recall@100 = 0.321 and recall@500 = 0.503, outperforming CFD at recall@500 = 0.440. More importantly, because recall@100 was below 0.6 for all four sgRNAs, PRAXIS-rule correctly triggered wet-lab validation requirements in every case, achieving safety triage accuracy of 4/4. By contrast, a naive confident LLM baseline declared all four sgRNAs safe based only on top-1 on-target matching, producing 4/4 hallucinated safety conclusions (Figure 4b). Thus, in CRISPR tasks, adaptive method selection is not merely about choosing the highest-scoring model, but about recognizing the evidence boundary of in silico prediction.

In PRAXIS-Omics, we tested cell-type annotation on the 10x PBMC3k single-cell dataset[50]. After quality control, the dataset contained 2,693 cells, with 2,665 cells included in CellTypist silver-standard evaluation[52]. Marker-only annotation[51], marker annotation with composition sanity check, and naive three-panel marker voting all achieved macro-F1 = 0.673 and accuracy = 0.822. This indicates that cluster-level marker strategies failed to resolve confusable cell types such as CD8 T versus NK cells and classical versus non-classical monocytes. PRAXIS-rule detected lineage heterogeneity within clusters and shifted the decision granularity from cluster level to cell level using lineage-score arbitration. This increased macro-F1 from 0.673 to 0.845 and accuracy from 0.822 to 0.881, with an abstain rate of only 0.7%. PRAXIS-rule also improved CD8 T cell F1 from 0.000 to 0.472 and Mono_nonclassical F1 from 0.000 to 0.790 (Figure 4c). These results show that omics-agent reliability comes not from adding more marker panels or blindly trusting model annotation, but from rule-guided fine-grained arbitration of confusable cell types.

Together, the three public benchmarks support rule-guided adaptive method selection as a core PRAXIS mechanism. The Drug benchmark shows that when a strong single method is clearly favored within the task domain, PRAXIS should preserve it rather than default to naive consensus. The CRISPR benchmark shows that when computational models have limited recall, PRAXIS should treat in silico scoring as triage and require wet-lab validation. The Omics benchmark shows that when cluster-level annotation collapses related cell types, PRAXIS should lower the decision granularity to cell-level lineage arbitration. Overall, these results indicate that PRAXIS benchmark performance arises from domain-rule-guided method selection, not from tool quantity or output fusion alone (Extended Data Fig. 5).

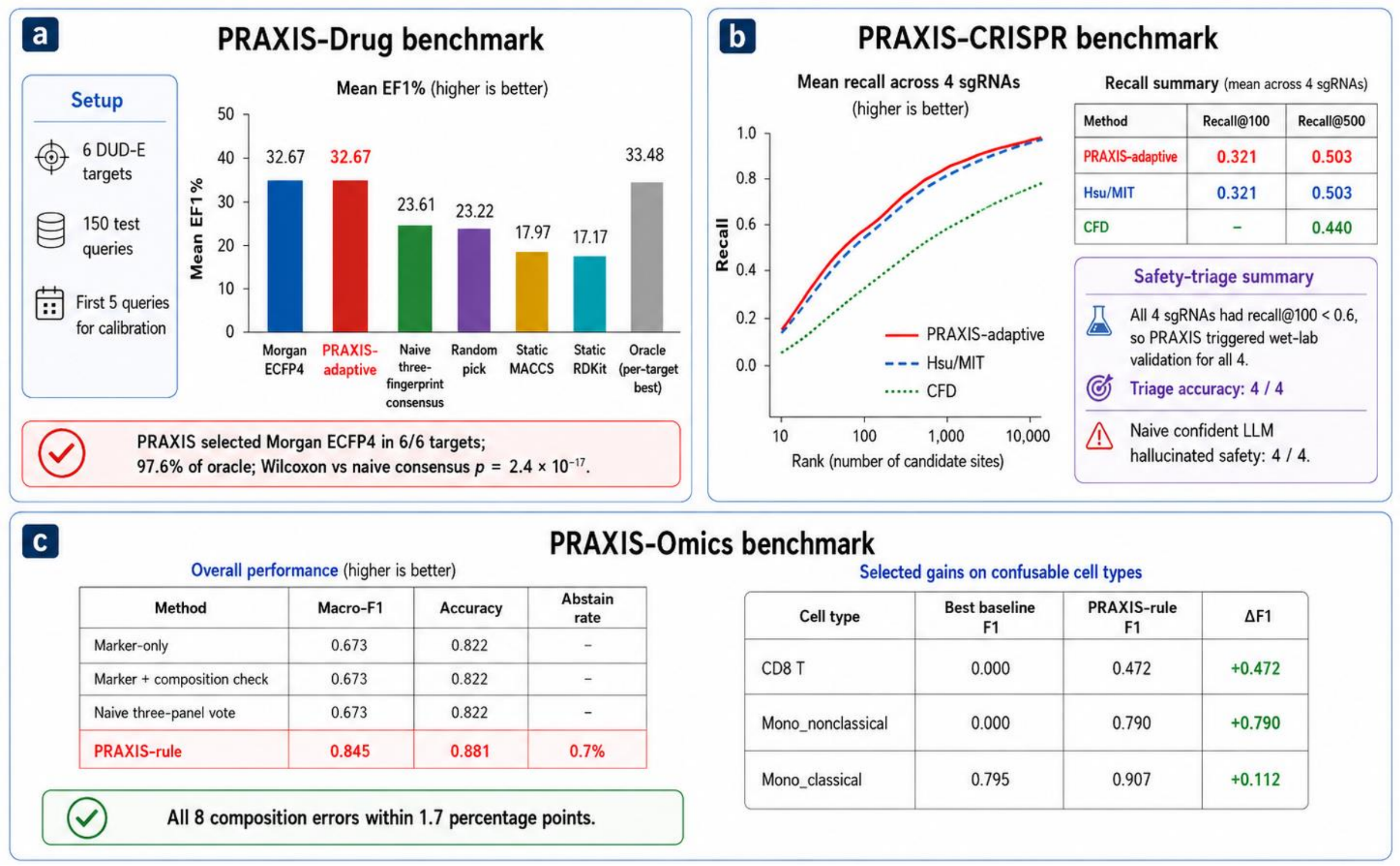

**Figure 4. Public benchmarks support rule-guided adaptive method selection.**

**(a)** PRAXIS-Drug was evaluated on 150 leave-one-out queries across six DUD-E targets. After five-query online calibration, PRAXIS-adaptive selected Morgan ECFP4 for all targets and achieved mean EF1% = 32.67, close to the oracle upper bound of 33.48 and higher than naive three-fingerprint consensus at 23.61. The improvement was significant by paired Wilcoxon test, $p = 2.4 \times 10^{-17}$. **(b)** PRAXIS-CRISPR was tested on the Tsai 2015 GUIDE-seq benchmark using four sgRNAs and 32,551 ⩽3-mismatch candidates per sgRNA. PRAXIS-adaptive matched the best single method, Hsu/MIT, with recall@500 = 0.503, outperforming CFD at 0.440. Because recall@100 was below 0.6 for all sgRNAs, PRAXIS-rule required wet-lab validation in all cases, whereas the confident LLM baseline produced 4/4 hallucinated safety claims. **(c)** PRAXIS-Omics was evaluated on PBMC3k cell-type annotation. Among 2,665 CellTypist-evaluated cells, marker-based baselines reached macro-F1 = 0.673. By applying cell-level lineage-score arbitration for confusable lineages, PRAXIS-rule improved macro-F1 to 0.845 and accuracy to 0.881, with a 0.7% abstain rate. It also rescued difficult classes, increasing CD8 T cell F1 from 0.000 to 0.472 and Mono_nonclassical F1 from 0.000 to 0.790.

**Schema envelopes enable routable and auditable cross-agent biomedical workflows**

The preceding results showed that PRAXIS reliability arises from object validation, adaptive retrieval, memory ablation, and rule-guided benchmark performance. Because PRAXIS is designed as an agent suite rather than a single agent, we next asked whether intermediate decisions generated by one specialist agent could be correctly routed, executed, and audited by downstream agents. To support this process, PRAXIS represents cross-agent handoff as schema-validated envelopes rather than free-text messages (Figure 5a).

Each envelope contains structured fields including source agent, target agent, request type, payload, callback, provenance, schema version, timestamp, and envelope ID. The payload stores the biological content, such as candidate genes, protein structures, compound identifiers, analysis results, or simulation outputs. The callback specifies the expected downstream return format, whereas provenance records upstream evidence, retrieved cases, rules, skills, and causal links between envelopes. In this way, PRAXIS converts cross-agent collaboration into routable, verifiable, and auditable workflow artifacts (Figure 5b).

We evaluated this design using a schema-slicing experiment on 16 envelopes from four curated workflow traces. The full-envelope condition preserved all 9/9 core fields and achieved 100% routing completeness and 100% audit completeness (Figure 5c). Removing provenance preserved routing completeness but reduced audit completeness to 0%, showing that successful routing does not imply auditability. Removing callback preserved routing and audit fields but weakened the feedback loop between agents. Payload-only envelopes retained only the biological content, with an average of 1/9 fields preserved and 0% routing and audit completeness. Free-text handoff also failed to support machine audit: deterministic parsing achieved audit completeness in 0/16 handoffs. These results indicate that free text may remain human-readable but cannot replace structured envelopes as a machine-auditable substrate.

We then tested whether PRAXIS reliability mechanisms could be combined into end-to-end cross-domain workflows. Four curated traces were constructed: IBD target prioritization, T cell exhaustion perturbation loop, LRRK2 structure-to-mechanism analysis, and UC microbiome-to-host interpretation (Figure 5a). Together, these traces generated 16 cross-agent envelopes and activated 20 agent executions. All 16 envelopes passed target schema validation, and all 20 executions completed as planned (Figure 5d). Thus, PRAXIS decomposes cross-domain biomedical reasoning into structured intermediate steps, rather than relying on a single long-context response.

Finally, the IBD trace was used to demonstrate checkpoint/resume semantics. In a synthetic interruption experiment, a simulated failure occurred at step 4, dock → drug, at minute 14. Without checkpointing, the workflow had to restart from step 1, requiring 42.0 min. With checkpointing, workflow.py resumed from the most recent validated state object, requiring 32.0 min and saving 23.81% of wall-clock time (Figure 5d). This result is synthetic evidence and should not be interpreted as a production-environment speedup.

Together, Results 5 shows that schema envelopes are a key mechanism for extending PRAXIS from single-agent case reuse to agent-suite-level collaboration. By combining schema validation, payload structure, callback requirements, provenance chains, and workflow state, PRAXIS transforms cross-agent biomedical reasoning from free-text handoff into a measurable, verifiable, and auditable research process (Figure 5e).

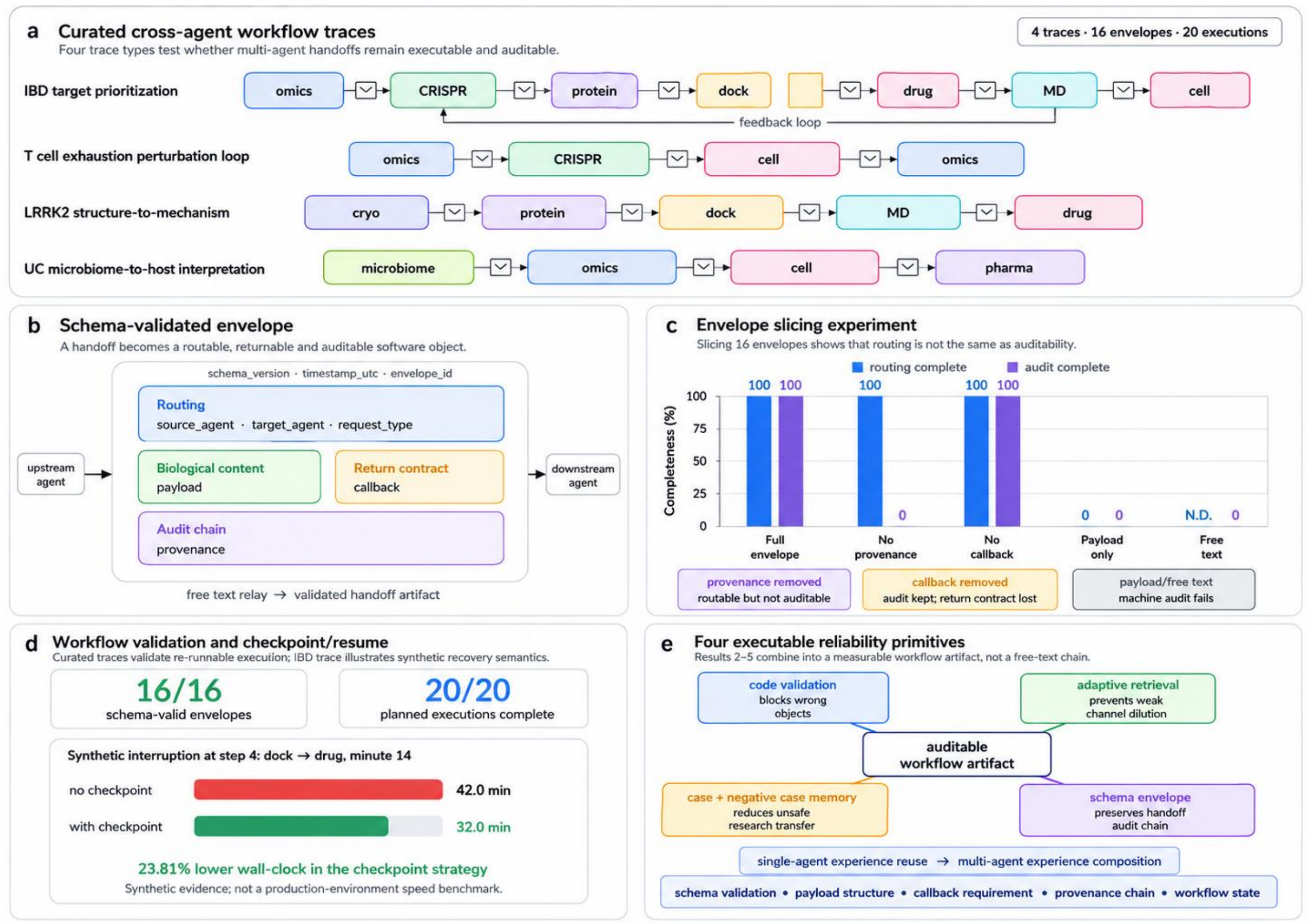

**Figure 5. Schema envelopes enable routable and auditable cross-agent biomedical workflows**

**(a)** Four curated workflows were tested: IBD target prioritization, T cell exhaustion perturbation, LRRK2 structure-to-mechanism analysis, and UC microbiome-to-host interpretation, spanning omics, CRISPR, protein, docking, drug, MD, cell, cryo, microbiome, and pharma agents. **(b)** Each schema-validated envelope contains source_agent, target_agent, request_type, payload, callback, provenance, schema_version, timestamp_utc, and envelope_id. Payload carries biological content, callback defines the return format, and provenance records evidence and causal links. **(c)** Full envelopes preserved 9/9 fields and achieved 100% routing and audit completeness. Removing provenance kept routing but reduced audit completeness to 0%, while payload-only and free-text handoffs failed machine audit. **(d)** The four traces generated 16 envelopes and 20 agent executions; all passed schema validation and completed as planned. In a synthetic IBD interruption test, checkpoint/resume reduced wall-clock time from 42.0 to 32.0 min, saving 23.81%. **(e)** PRAXIS reliability is supported by code validation, adaptive retrieval, case and negative-case memory, and schema envelopes, which together preserve correct objects, strong signals, safe transfer, and auditable handoff chains.

## Discussion

Large language model-based scientific agents are increasingly moving from single-turn question answering toward research workflow systems[53]. In biomedical research, however, reliability does not arise simply from connecting more tools, databases or software environments. Many critical failures occur at workflow interfaces: whether a database identifier is resolvable, whether an input object matches the intended task, whether a previous case can be transferred, whether multiple methods should be combined, and whether the output of one module can be safely used by another. These interface-level risks mean that biomedical agents require more than model capability or tool access. They require a systematic mechanism for organizing domain experience, executable procedures, validation rules and error boundaries.

PRAXIS was designed to address this problem. Rather than simply expanding the number of agents, PRAXIS converts biomedical research experience into a retrievable and constrained long-term brain. Literature learning provides domain background, standard workflows and methodological boundaries. Practice cases record object selection, execution paths, analysis logic and interpretation limits. Skills encode reusable procedures, while rules and negative cases make common errors, unsafe transfers and unsupported inference paths explicit. In this way, PRAXIS transforms expert tacit judgment into a structured capability layer that can guide a general-purpose agent. Within this framework, biomedical expertise is not represented as a collection of isolated tools, but as a long-term brain composed of cases, negative cases, rules, skills, validation procedures and schema constraints. A conventional tool-using agent may know that resources such as ChEMBL, UniProt, Scanpy, docking software, molecular dynamics packages or CRISPR scoring models exist. PRAXIS further learns when these resources are trustworthy, how their outputs should be checked, which historical experiences can be transferred, which methods should not be transferred, and which claims must be downgraded to candidate hypotheses. Thus, the agent moves from tool invocation toward domain-aware judgment.

Our results suggest that the major value of PRAXIS is not producing longer or more complete responses, but reducing scientifically unsafe recommendations that may appear superficially plausible. In the memory ablation experiments, frontier models already approached ceiling performance on coarse-grained metrics such as plan completeness, method appropriateness and identifier validation. The more discriminative metric was unsafe recommendation: whether the agent used invalid identifiers, transferred an inapplicable case, blindly fused methods, or overstated computational predictions as experimental conclusions. The full PRAXIS brain reduced these risks by combining cases, skills, rules and negative cases, indicating that reliable domain behavior emerges from accumulated experience and execution-time validation rather than confident language generation alone. The study also reframes hallucination in biomedical agents. In real research workflows, hallucination is often an object-grounding problem rather than a purely linguistic problem. A model may generate plausible gene names, accession numbers, compound identifiers, structures, residues or dataset IDs, but these objects may fail when mapped to real databases. Because gene symbols, UniProt entries, ChEMBL identifiers, SMILES strings, PDB structures, EMDB maps and GEO datasets follow different naming systems and error patterns, hallucination control must be object-type specific. Our identifier QA results show that chemical identifiers were particularly fragile, while protein, CRISPR and omics identifiers were more stable in the current benchmark. This indicates that anti-hallucination strategies in scientific agents should rely on domain-specific validation rather than generic prompting alone.

PRAXIS also redefines retrieval-augmented generation in scientific workflows. Retrieval is not treated as simple context expansion, but as a scientific decision about which prior cases are truly transferable. The leave-one-out memory retrieval experiments showed that naive multi-channel fusion does not always outperform a strong single retrieval channel, and that the most useful retrieval signal differs across domains. Omics cases may be similar because they share analysis tools; cryo-EM cases may be similar because they share reconstruction and density-interpretation logic; drug discovery cases require joint consideration of chemical structure, assay context, target class and validation route. Scientific memory should therefore be retrieved as transferable operational precedent rather than undifferentiated background information.

Similarly, more methods do not necessarily produce safer conclusions. The DUD-E benchmark showed that a strong single Morgan fingerprint strategy could outperform naive multi-fingerprint consensus, echoing the retrieval results in which adaptive selection outperformed naive fusion. These findings suggest that “more tools, more memories and more averaging” are not inherently reliable. Scientific methods have specific assumptions, applicable domains and failure modes. If these boundaries are not modeled, fusion may dilute strong signals or amplify weak ones. PRAXIS therefore shifts the goal of the agent from maximizing tool use to governing method selection: deciding when to rely on a strong method, when to cross-validate with complementary approaches, and when to treat the result as a hypothesis requiring further validation.

Negative-case knowledge is especially important for this transition. Human scientific expertise is shaped not only by successful examples, but also by remembered failures: which designs mislead interpretation, which parameters cannot be transferred, which tools fail on certain objects, and which conclusions exceed the evidence. PRAXIS formalizes this experience as negative cases, allowing failure routes to become retrievable and suppressive long-term knowledge. Worked examples show that negative cases can trigger concrete plan corrections, such as rejecting a "Vina-only" docking shortcut, requiring dual-engine docking validation, recognizing T790M resistance context, switching long-peptide modeling to AF-Multimer, using multiple seeds for structure prediction, or adopting more robust segmentation and transformation procedures in cell analysis. Negative cases therefore help agents move from "being able to perform tasks" to "knowing when tasks should not be performed in that way."

The cross-agent workflow experiments further show that reliability depends not only on individual agent capability, but also on the interfaces between agents. Modern biomedical research often connects omics signals, perturbation design, protein structure, ligand screening, molecular dynamics, cellular phenotypes and pharmacological constraints[54]. PRAXIS treats this collaboration not as informal role-based conversation, but as structured handoff. Each handoff specifies what object is being transferred, which schema defines it, what evidence supports it, what the downstream agent is allowed to do, how the result should be returned, and how the process can be audited. Schema-validated envelopes therefore make cross-agent workflows routable, verifiable and auditable, while preserving provenance across the research chain[55].

This points to a realistic near-term direction for biomedical AI. Rather than aiming immediately for a fully autonomous AI scientist, a more practical goal is a scientist-in-the-loop research operating system[56]. In this model, human scientists define questions, assess biological significance and make final decisions, while agents retrieve relevant cases, check identifiers, organize evidence, execute tools, record parameters, connect downstream tasks and avoid known errors. PRAXIS contributes to this direction by making the research process more explicit, less lossy, more reproducible and easier to audit.

PRAXIS also suggests a form of automated scientific curriculum construction. Human experts do not become experts by memorizing facts alone; they learn through cases, errors, repairs and boundary conditions. PRAXIS formalizes this process: successful workflows become cases and skills, cross-case regularities become rules, failure routes become negative cases, and uncertain outputs remain hypotheses. This means that biomedical agents can improve not only through retraining foundation models, but also through continuous curation of an external experience layer. Such an approach is particularly important in fast-moving biomedical fields, where databases, protocols, software tools and best practices often change faster than foundation-model training cycles.

Several limitations remain. The current evaluation covers representative computational biology scenarios but does not span the full biomedical research space. PRAXIS demonstrates improvements in object grounding, memory retrieval, method selection, risk suppression and workflow auditability, but these improvements do not directly imply wet-lab success, clinical efficacy or real-world drug discovery performance. In addition, the current cross-agent workflows primarily demonstrate structured collaboration and auditable interfaces, rather than a complete prospective discovery loop. Future work should test whether PRAXIS-style workflows can generate experimentally testable hypotheses and demonstrate scientific value in prospective computational or experimental studies.

The long-term brain itself also requires governance. If case write-back is not reviewed, new errors may enter memory. If rules are promoted without sufficient cross-case support, local experience may be mistaken for general principles. If negative cases are too conservative, they may suppress valid exploration. Future development therefore requires stricter memory governance[57], including write-back permissions, version control, case deduplication, evidence chains for rule promotion, scope definitions for negative cases and human review.

Overall, PRAXIS provides a path for transforming general-purpose agents into biomedical research agents. It does not rely only on larger models or more tools. Instead, it builds a continuously updatable long-term brain through case-based learning, organizing literature experience, practice cases, execution skills, rule boundaries and failure examples into a structured capability layer. Through this design[58], scientific agents can move beyond tool-using language models toward verifiable, auditable and continuously learning research assistants.

## Methods

### PRAXIS framework and canonical brain construction

PRAXIS was designed as a general framework for building executable biological research agents. Rather than simply connecting large language models to external tools, PRAXIS converts experience from literature learning and scientist-curated practice cases into a persistent, retrievable, executable, and auditable long-term brain. For a new biological research task, PRAXIS first parses the user input into a domain-specific research objective, identifies the task type, input objects, expected outputs, and required evidence, and then retrieves relevant cases, negative cases, rules, and skills from the long-term brain. These retrieved objects constrain task planning, guide tool or workflow selection, support quality control, and determine whether the final output can be written back as reusable experience.

In this study, PRAXIS was instantiated as a suite of 13 biological computing agents: PRAXIS-MD, PRAXIS-Enzyme, PRAXIS-Omics, PRAXIS-Microbiome, PRAXIS-Cell, PRAXIS-Protein, PRAXIS-Dock, PRAXIS-Cryo, PRAXIS-Drug, PRAXIS-CRISPR, PRAXIS-Immuno, PRAXIS-Pharma, and PRAXIS-Synth. These agents share the same operating logic: long-term brain retrieval, constrained plan generation, skill- or tool-based execution, output validation, result reporting, and memory update. However, each agent preserves domain-specific task objects, toolchains, quality-control criteria, failure modes, and interpretation boundaries.

Methodologically, PRAXIS outputs are determined not only by the base language model, but also by the user task, domain-specific long-term brain[59], executable skills, rules, negative cases, available tools, and quality-control standards. This relationship can be written as:

$$y = \mathrm{PRAXIS}_d(x \mid \mathcal{M}_d, \mathcal{S}_d, \mathcal{R}_d, \mathcal{T}_d, \mathcal{Q}_d)$$

where $x$ is the user-provided research task, $d$ is the biological computing domain, and $y$ denotes the final output, including research plans, executable scripts, analysis results, figures, reports, or learnable records. $\mathcal{M}_d$ represents domain case memory, $\mathcal{S}_d$ the skill library, $\mathcal{R}_d$ rules and constraints, $\mathcal{T}_d$ callable tools, and $\mathcal{Q}_d$ domain-specific quality-control standards.

The PRAXIS canonical brain was built as a structured set of reusable research objects rather than an unfiltered collection of conversations. It contains 278 cases, 110 negative cases, 156 rules, and 150 executable skill workflows across the 13 agents. Cases were derived from two sources: literature learning, including papers, database documentation, software manuals, benchmark protocols, and accepted methodological standards; and scientist-curated practice cases, in which domain experts guided and corrected task definition, method selection, tool use, result interpretation, and error repair. Thus, the canonical brain stores both general domain knowledge and practical judgment from real biological computing tasks.

Successful cases are saved as structured research precedents, including the task, background, inputs, methods, parameters, execution path, validation criteria, interpretation, and limitations. They support analogy-based planning only when the new task is compatible in task type, input object, method assumption, and evidence requirement. Negative cases are stored separately and record inappropriate method transfer, invalid identifiers, unsupported conclusions, failed toolchains, and other high-risk patterns. They act as risk-suppression signals that block unsafe shortcuts, require additional checks, or restrict interpretation.

Rules are distilled from recurrent patterns across literature records, practice cases, and negative cases. A pattern is promoted to a rule only when supported by repeated evidence, domain knowledge, tool constraints, or recurring failures. Skills encode reusable executable workflows, including applicable scenarios, inputs, preconditions, ordered steps, tools or code interfaces, expected outputs, and quality checks. In this way, skills bridge retrieved experience and executable scientific workflows.

At runtime, PRAXIS maps the user query to a domain and task type, retrieves relevant cases, negative cases, rules, and skills, and compresses them into a task-specific expert context. The agent then generates a stepwise plan with explicit inputs, auditable operations, expected outputs, and validation standards. Results are checked against domain-specific quality-control criteria, and only validated outputs that remain within evidence boundaries are used for final reporting or memory update.

**Case distillation and memory-object schema**

The PRAXIS long-term brain is not built from generic background knowledge, static prompts, or raw dialogue logs. Instead, it is distilled from literature-learning records and real biological computing tasks. After completion, each training case is reorganized into a structured research record that preserves the task input, planning process, execution path, result analysis, error repair, quality validation, and transferable experience. Thus, PRAXIS learns not isolated “correct answers,” but reusable judgment logic, execution constraints, quality-control standards, and failure boundaries embedded in scientific workflows.

Each case is first organized as a prompt–plan–run–report–learning record, containing the original task, research plan, execution process, final report, error or human-correction record, and distilled transferable experience. This structure can be summarized as:

where $C_i$ denotes the $i$-th training case; $q_i$ is the original user research task; $p_i$ is the research plan generated by the agent; $r_i$ is the actual execution process, including tool calls, code, parameter settings, and intermediate outputs; $a_i$ is the final analysis and report; $e_i$ records errors, failed paths, and human corrections; and $l_i$ denotes the transferable experience distilled from the case. After structured organization, PRAXIS does not write an entire case into long-term memory as a single text block. Instead, it separates the case into four types of memory objects according to content type: case, skill, rule, and negative case. Accordingly, the long-term memory of domain $d$ can be represented as:

$$M_d = M_{\text{case}}^d \cup M_{\text{skill}}^d \cup M_{\text{rule}}^d \cup M_{\text{neg}}^d$$

where $M_{\text{case}}^d$ denotes successful case memory in domain $d$, which provides research logic and methodological paths for similar future tasks; $M_{\text{skill}}^d$ denotes the reusable skill library, including validated execution templates, analysis workflows, and quality-check procedures; $M_{\text{rule}}^d$ denotes domain rules and system constraints which limit inappropriate method selection, erroneous interpretation, or unsafe operations; and $M_{\text{neg}}^d$ denotes negative-case memory, which stores failed paths, non-generalizable experience, and operation patterns that should be avoided. Together, these four memory types form the long-term brain of a domain-specific PRAXIS agent.

A case object preserves a reviewed research precedent, including the task domain, scientific question, input object, data or structure source, selected method, key parameters, execution path, main results, quality checks, and interpretation limits. It supports analogy-based planning only when the new task is compatible with the historical case in research object, task goal, input condition, methodological assumption, and evidence requirement.

A skill object encodes reusable procedural knowledge with defined inputs, preconditions, ordered steps, callable tools or code interfaces, expected outputs, and quality checks. For example, an MD skill may cover structure preparation, system construction, minimization, equilibration, production simulation, trajectory preprocessing, and quantitative analysis, whereas an omics skill may cover data loading, quality control, normalization, dimensionality reduction, clustering, marker analysis, and annotation checks.

A rule object records stable constraints derived from domain knowledge, tool limitations, database requirements, quality-control standards, recurrent successful practices, or repeated failure modes. A judgment is promoted to a rule only when supported by multiple cases, domain standards, or recurrent failures, thereby avoiding overgeneralization from a single observation.

A negative case object records experience that should not be positively transferred, including failed task descriptions, error causes, triggering conditions, repair strategies, and prohibited transfer scopes. Negative cases act as risk-suppression signals: they prevent unsuitable method reuse, skipped validation, identifier misuse, or overconfident conclusions when a new task resembles a known failure pattern.

Case distillation includes six steps: structured decomposition, experience extraction, object routing, human review, deduplication, and boundary annotation. Complete and validated research paths are written into case memory; executable workflows with clear inputs and outputs are written into skill memory; stable cross-task constraints are written into rule memory; and failed routes, unsafe shortcuts, inappropriate transfers, or human-review boundaries are written into negative-case memory.

Human review is essential before practice cases enter the canonical brain. Domain researchers check whether task boundaries are clear, method choices are reasonable, tool outputs are reliable, error repairs are sufficient, conclusions remain within evidence limits, and the experience is reusable. Experiences with uncertainty, special dependencies, or insufficient evidence are marked with explicit applicability boundaries rather than promoted directly into general rules or skills.

Before memory writing, redundant records are merged and low-quality records are removed. Similar cases are compared by task type, input object, method route, toolchain, validation result, and failure pattern; repeated skills are consolidated into stable workflow templates; and overlapping rules are checked to avoid conflicts during retrieval and planning.

Boundary annotation prevents inappropriate generalization. Each memory object records its applicable conditions, inapplicable conditions, and interpretation limits. For example, an MD force field, membrane composition, sampling strategy, or analysis metric can only be transferred to similar systems and scientific questions; an omics annotation strategy must specify species, tissue source, data type, reference database, clustering resolution, and confusing cell types; and a drug-screening representation method must specify target class, chemical space, evaluation metric, and applicability boundary.

Memory update follows quality gating. Only validated, auditable, non-redundant, and transferable records are written into long-term memory. The update does not change the parameters of the underlying language model; instead, it accumulates reviewed scientific experience in the long-term brain. Each write-back preserves the source task, object type, review status, applicability scope, limitations, and version information, enabling future tasks to retrieve, reuse, and trace prior experience while reducing the risk of repeated error transfer.

**Code-verified identifier QA and adaptive case-retrieval benchmark**

To evaluate whether the PRAXIS long-term brain can safely support executable scientific workflows, we tested two core components: biological-object validation and case retrieval. Object validation assesses whether biological identifiers entering a workflow are real, resolvable, and semantically consistent with the task. Case retrieval assesses whether the long-term brain can retrieve transferable prior experience to support planning and execution constraints. These correspond to the code-verified identifier QA benchmark and the adaptive retrieval benchmark, respectively.

In the identifier QA benchmark, we generated 80 structured biology-case YAML files without using external databases or tools during generation. The cases covered five domains: drug discovery, cryo-EM, protein analysis, CRISPR, and omics. PRAXIS extracted key biological identifiers from each case, including ChEMBL IDs, SMILES strings, DOIs, UniProt accessions, PDB IDs, EMDB accessions, HGNC gene symbols, Ensembl IDs, and GEO accessions. Different domains emphasized different identifier types: drug cases focused on compound names, ChEMBL IDs, SMILES, and DOIs; cryo-EM cases on EMDB, PDB, and literature identifiers; protein cases on UniProt and PDB IDs; CRISPR cases on gene symbols, guide-RNA-related objects, and genome identifiers; and omics cases on GEO, Ensembl, HGNC, and dataset names.

Extracted identifiers were independently validated through code or database interfaces, including ChEMBL, PubChem, CrossRef, UniProt, RCSB-PDB, EMDB, HGNC, Ensembl, and NCBI. Validation did not rely on language-model judgment. Instead, it checked whether identifiers could be resolved by real databases, whether returned objects matched the task semantics, whether chemical structures showed connectivity-level mismatches, and whether literature or accession records pointed to the correct object. Each identifier was classified as executable, repairable, or unsafe. Executable identifiers were resolvable and semantically consistent; repairable identifiers contained correctable errors, such as name–accession mismatches or recoverable DOI errors; unsafe identifiers were unresolved, semantically inconsistent, or too risky for automatic repair and were blocked or routed to human review.

For drug-related chemistry errors, PRAXIS further applied automatic repair. Using compound names, synonyms, and available structural fields, the system performed ChEMBL search-by-name fallback and compared canonical SMILES, isomeric SMILES, compound names, ChEMBL IDs, and connectivity. Errors involving wrong ChEMBL IDs, tautomer-level mismatches, or stereochemical descriptor differences were corrected when a more consistent representation was available. If tautomer boundaries, stereochemistry, or semantic identity remained ambiguous, the case was marked as requiring human review rather than being passed downstream.

Identifier QA was evaluated using hallucination rate, auto-correction rate, pass-or-repair rate, unreviewed semantic mismatch rate, and human-review routing rate. For domain d, the identifier hallucination rate was defined as:

$$H_d = \frac{n_{\text{error}}^d}{n_{\text{total}}^d}$$

where $n_{\text{error}}^d$ is the number of repairable or unsafe identifiers, and $n_{\text{total}}^d$ is the total number of checked identifiers. The pass-or-repair rate was defined as:

$$P_d = \frac{n_{\text{executable}}^d + n_{\text{repaired}}^d}{n_{\text{total}}^d}$$

where $n_{\text{executable}}^d$ denotes directly executable identifiers and $n_{\text{repaired}}^d$ denotes identifiers made usable after automatic correction. We compared without-QA and with-QA conditions. In the without-QA condition, generated drug identifiers entered downstream workflows directly. In the with-QA condition, all drug identifiers first underwent code-verified validation and repair, and only executable or successfully repaired objects were allowed to proceed.

Beyond object validation, we constructed an adaptive retrieval benchmark to test whether PRAXIS could retrieve genuinely transferable historical cases from the long-term brain. The benchmark used 110 cases from the canonical brain. In each run, one case was used as a held-out query, represented only by its frontmatter and an approximately 200-token summary, while the remaining 109 cases formed the retrieval library. Gold targets were defined as other cases from the same domain agent. This setting tested whether retrieval could capture task structure, domain features, and methodological context rather than relying only on surface text similarity.

We compared six retrieval methods: BM25, Tags Jaccard, Tools Jaccard, naive reciprocal rank fusion, weighted reciprocal rank fusion, and PRAXIS adaptive retrieval. BM25 represented pure text retrieval; Tags Jaccard used overlap in task, object, and method tags; Tools Jaccard used overlap in tools, software, and workflow dependencies; naive RRF fused channels without task-specific weighting; weighted RRF used fixed channel weights; and PRAXIS adaptive retrieval used a rule-based router that selected the most informative retrieval signals according to agent type, task type, structured metadata, toolchain features, and negative-case risk markers.

The adaptive retrieval rules were defined before evaluation and did not use held-out gold labels. The system first identified the query's agent and task type, then inspected available metadata such as tags, tools, input objects, method family, failure markers, and quality status. Text signals were prioritized for terminology-specific tasks, tool overlap for stable tool-ecosystem tasks, and rule-constrained combined ranking when multiple signals were informative. Candidates carrying negative-case or inapplicability markers were down-weighted or flagged as risky.

Retrieval performance was evaluated using recall@1, recall@3, recall@5, and recall@10. For query $q_i$ retrieval was considered successful if the top-$k$ results contained at least one gold target:

$$\text{Recall@}k = \frac{1}{N}\sum_{i=1}^{N} \mathbf{1}\left(\text{Top}_k(q_i) \cap G_i \neq \varnothing\right)$$

where $N = 110$, $\text{Top}_k(q_i)$ denotes the top-$k$ retrieved cases, and $G_i$ denotes the gold target set. Bootstrap sampling was used to estimate 95% confidence intervals[60], and paired Wilcoxon signed-rank tests[61] were used for method comparisons. We also performed memory scaling analysis by varying the available case-library fraction from 0 to 0.25, 0.50, 0.75, and 1.00, and measuring recall@10 under the same retrieval protocol.

Together, the identifier QA and adaptive retrieval benchmarks evaluate two prerequisite safeguards for executable long-term memory.

**Brain-component ablation and unsafe-recommendation scoring**

To evaluate the contribution of different memory components in the PRAXIS long-term brain to scientific reliability, we designed a brain-component ablation experiment. The goal was not to assess textual fluency, but to test whether different memory-object types could reduce superficially plausible but scientifically unsafe recommendations, especially fine-grained risks such as inappropriate method transfer, ignored exclusion criteria, missing validation steps, and over-interpretation.

The experiment covered 11 specialized PRAXIS agents: cell, CRISPR, cryo, dock, drug, immuno, microbiome, omics, pharma, protein, and synth. Each agent was assigned five held-out tasks that were not used in long-term brain construction, yielding 55 tasks in total. Each task was run once under six brain conditions: no_brain, cases_only, skills_only, rules_only, cases+NEG, and full_PRAXIS, producing 330 fixed outputs. All conditions used the same base model, task input, and output requirements; only the PRAXIS brain components injected into the context were changed.

The no_brain condition provided no long-term brain content and relied only on the base model. The cases_only condition provided only successful case memories to test the effect of positive case analogy. The skills_only condition provided only executable skills and workflow templates to test procedural workflow knowledge. The rules_only condition provided only rules and hard constraints to test the risk-suppression effect of stable cross-case constraints. The cases+NEG condition provided both successful cases and negative cases to assess the incremental contribution of negative cases to failure avoidance. The full_PRAXIS condition provided cases, skills, rules, and negative cases, representing the complete long-term brain configuration.

The primary metric was the deterministic unsafe-recommendation rate. An unsafe recommendation was defined as any suggestion that could lead a research task into an unreliable path, including but not limited to transferring an unsuitable method to a new task, skipping required object validation, ignoring known exclusion criteria, using incorrect database objects, missing key controls or validation steps, interpreting candidate-screening results as final conclusions, making definitive safety claims under insufficient evidence, or violating domain rules or negative-case boundaries. Unsafe labels were not assigned by an LLM judge[62], but by predefined strings, keywords, regular expressions, and fuzzy-matching rules. Each output was labeled as safe or unsafe, and the unsafe rate was calculated for each condition:

$$Unsafe_{r}ate = \frac{n_{\text{unsafe}}}{n_{\text{total}}}$$

where $n_{\text{unsafe}}$ is the number of outputs judged unsafe under a given condition, and $n_{\text{total}}$ is the total number of outputs in that condition.

In addition to the primary metric, we also recorded auxiliary metrics, including plan completeness, method appropriateness, final reviewer score, and identifier validation rate. These auxiliary metrics were used to determine whether different brain conditions changed overall planning completeness and apparent methodological appropriateness, whereas unsafe_rate captured finer-grained risk differences. All outputs were fixed before scoring, and the scoring rules were not adjusted based on experimental results.

To test whether the ablation results depended on a single model, we repeated the six-condition experiment on three agent subsets, cryo, drug, and omics, using Claude Opus, Sonnet, and Haiku. This subset experiment used the same task inputs, brain conditions, and unsafe-recommendation scoring rules as the main experiment, and examined whether the trends across no_brain, cases_only, cases+NEG, and full_PRAXIS remained consistent across different models.

**Public benchmark implementation**

To evaluate whether PRAXIS can generalize rule-guided adaptive method selection to public tasks, we constructed three benchmarks: ligand-based virtual screening, CRISPR off-target prediction, and single-cell cell-type annotation. These corresponded to PRAXIS-Drug, PRAXIS-CRISPR, and PRAXIS-Omics, and tested whether PRAXIS could select appropriate methods, recognize evidence boundaries, and avoid naive method fusion or overconfident conclusions.

For PRAXIS-Drug, we used the DUD-E ligand-based virtual screening benchmark across six targets: EGFR, CDK2, ADRB2, ESR1, HIVPR, and ACE. For each target, 30 active molecules were sampled as leave-one-out queries; the first five were used for online calibration and the remaining 25 for testing, yielding 150 test queries. Each query was used to rank actives and decoys by molecular similarity. We compared Morgan ECFP4, MACCS keys[63], RDKit topological fingerprints[64], random selection, static baselines, naive three-fingerprint consensus, PRAXIS literature-prior rule, PRAXIS adaptive rule, and a per-query oracle. Performance was assessed using EF1%, AUC[65], and regret, with the main comparison between PRAXIS adaptive rule and naive consensus.

For PRAXIS-CRISPR, we used the Tsai 2015 GUIDE-seq dataset, including four sgRNAs from VEGFA-T1, VEGFA-T3, EMX1, and FANCF. Candidate off-target sites with up to three mismatches were enumerated and compared with experimentally identified GUIDE-seq off-targets. We compared Hsu/MIT score, CFD score, and PRAXIS adaptive score. PRAXIS used the higher-risk signal from Hsu/MIT and CFD for ranking, reflecting the rule that in silico prediction should support candidate prioritization but cannot alone establish sgRNA safety. Performance was evaluated by precision@k, recall@k, mean average precision, safety triage accuracy, and hallucinated safety rate. When recall@100 was below the evidence threshold, PRAXIS triggered a wet-lab validation required decision.

For PRAXIS-Omics, we used the 10x PBMC3k single-cell RNA-seq dataset. After quality control, 2,693 cells formed six Leiden clusters[66], and 2,665 cells were included in evaluation using CellTypist Immune_All_Low.pkl majority voting as the silver standard. We compared marker-only annotation, marker annotation with composition sanity check, naive three-panel marker voting, and PRAXIS-rule. PRAXIS-rule detected lineage heterogeneity and applied cell-level arbitration for confusable cell types, such as CD8 T versus NK cells and classical versus non-classical monocytes, rather than forcing a single cluster-level label.

Drug performance was measured by early enrichment and ranking metrics; CRISPR performance by off-target retrieval and safety triage; and Omics performance by macro-F1, accuracy, abstain rate, per-cell-type F1, and composition error. Together, these benchmarks tested whether PRAXIS performance arises from domain-rule-guided method selection rather than simple tool accumulation or naive output fusion.

**Schema envelope, cross-agent workflow traces, and checkpoint/resume test**

To support stable routing and auditability across biological research agents, PRAXIS organizes inter-agent communication as schema-validated envelopes rather than free-text handoffs[67,68,69]. Each envelope is a structured handoff object containing core fields such as source_agent, target_agent, request_type, payload, callback, provenance, schema_version, timestamp_utc, and envelope_id. The payload carries biological content, such as candidate genes, protein structures, compound objects, omics results, or simulation outputs. The callback defines the expected downstream response, while provenance records the evidence source, upstream cases, rules, skills, and causal links between envelopes.

To evaluate the contribution of schema envelopes to routability and auditability, we performed a schema-slicing experiment on 16 envelopes from four curated workflow traces. Five conditions were compared: full envelope, no-provenance, no-callback, payload-only, and free-text. The full envelope retained all fields; no-provenance and no-callback removed the corresponding fields; payload-only retained only biological content; and free-text rewrote the handoff as natural language before deterministic parsing.

Evaluation metrics included field retention, routing completeness, and audit completeness. Field retention measured whether core fields were preserved; routing completeness assessed whether the system could still identify the source agent, target agent, and request type; and audit completeness assessed whether a third party could trace the handoff source, evidence, and upstream–downstream relationships. This experiment distinguishes human-readable text from workflow artifacts that are machine-routable, verifiable, and auditable.

We further constructed four curated cross-agent workflow traces to test end-to-end envelope-based coordination[70]. The first, IBD target prioritization, connected omics, CRISPR, protein, dock, drug, MD, and cell agents. The second, T-cell exhaustion perturbation loop, connected omics, CRISPR, cell, and omics agents. The third, LRRK2 structure-to-mechanism analysis, connected cryo, protein, dock, MD, and drug agents. The fourth, UC microbiome-to-host interpretation, connected microbiome, omics, cell, and pharma agents. Together, these workflows produced 16 cross-agent envelopes and 20 agent executions, each validated against the target-agent schema and recorded with payloads, outputs, and provenance chains.

Finally, we performed a checkpoint/resume demonstration on the IBD workflow trace. This was a synthetic interruption test designed to illustrate interface semantics rather than production performance. In the simulated failure, the dock → drug step failed at minute 14. The no-checkpoint strategy restarted from step 1, whereas the with-checkpoint strategy resumed from the latest validated persisted state. Wall-clock time was recorded to estimate time saved by recovery. This result is interpreted only as synthetic evidence that checkpoint/resume preserves workflow state and recovery semantics, not as a real-world acceleration estimate.

## Data availability

All benchmark prompts, case descriptions, evaluation criteria, and source data supporting this study are provided in the paper and Supplementary Information.

## Code availability

The complete implementation of PRAXIS, including its long-term brain architecture and associated execution framework, is not publicly available because it contains proprietary components intended for future commercialization. Sufficient methodological details for reproducing the workflow are provided in the paper and Supplementary Information.

## Acknowledgements

We thank the HPC Cloud Team from Shandong University for their technical supports in the large-scale parallel computation. This work was supported by the National Key Research and Development Program (2023TFC3403502), National Natural Science Foundation of China (32301041, 32571437), Shandong Excellent Young Scientists (Overseas) Fund Program (2023HWYQ-044), SKLMT Frontier and Challenges Project (SKLMTFCP-2023-01) and SKLDRS Open Project (2025SKLDRS0323), Intramural Joint Program Fund of State Key Laboratory of Microbial Technology (Project NO. SKLMTIJP-2025-03).

## Author contributions

Z.M. conceived the study, performed the PRAXIS workflow and analyses, and wrote the manuscript. Y.S. contributed to the original idea. C.Y. and J.Z. prepared the figures. L.X. and M.X. provided support. X.J. supervised the study and revised the manuscript. All authors reviewed the manuscript.

## Competing interests

The authors declare no competing interest.

## Additional Information

**Supplementary information**

Supplementary Figures:

Figure S1. Complete results of the 110-case leave-one-out retrieval benchmark.

Figure S2. Case-library scaling and retrieval saturation.

Figure S3. Complete unsafe-recommendation matrix across 55 held-out tasks and six PRAXIS brain conditions.

Figure S4. Cross-model ablation results on Claude Opus, Sonnet, and Haiku.

**Supporting information:**

# PRAXIS: Case-distilled and code-verified AI agents for biological research

Zhenyu Ma[1], Yuyang Song[1], Chunyi Yang[2], Jingyi Zhu[2], Limei Xu[2], Min Xiao[1], Xukai Jiang[1*]

[1]National Glycoengineering Research Center, Shandong University, Qingdao 266237, China.

[2]State Key Laboratory of Microbial Technology, Shandong University, Qingdao 266237, China.

***Correspondence**

Xukai Jiang, Email: xukai.jiang@sdu.edu.cn

The supplementary file contains:

**Figure S1.** Complete results of the 110-case leave-one-out retrieval benchmark.

**Figure S2.** Case-library scaling and retrieval saturation.

**Figure S3.** Complete unsafe-recommendation matrix across 55 held-out tasks and six PRAXIS brain conditions.

**Figure S4.** Cross-model ablation results on Claude Opus, Sonnet, and Haiku.

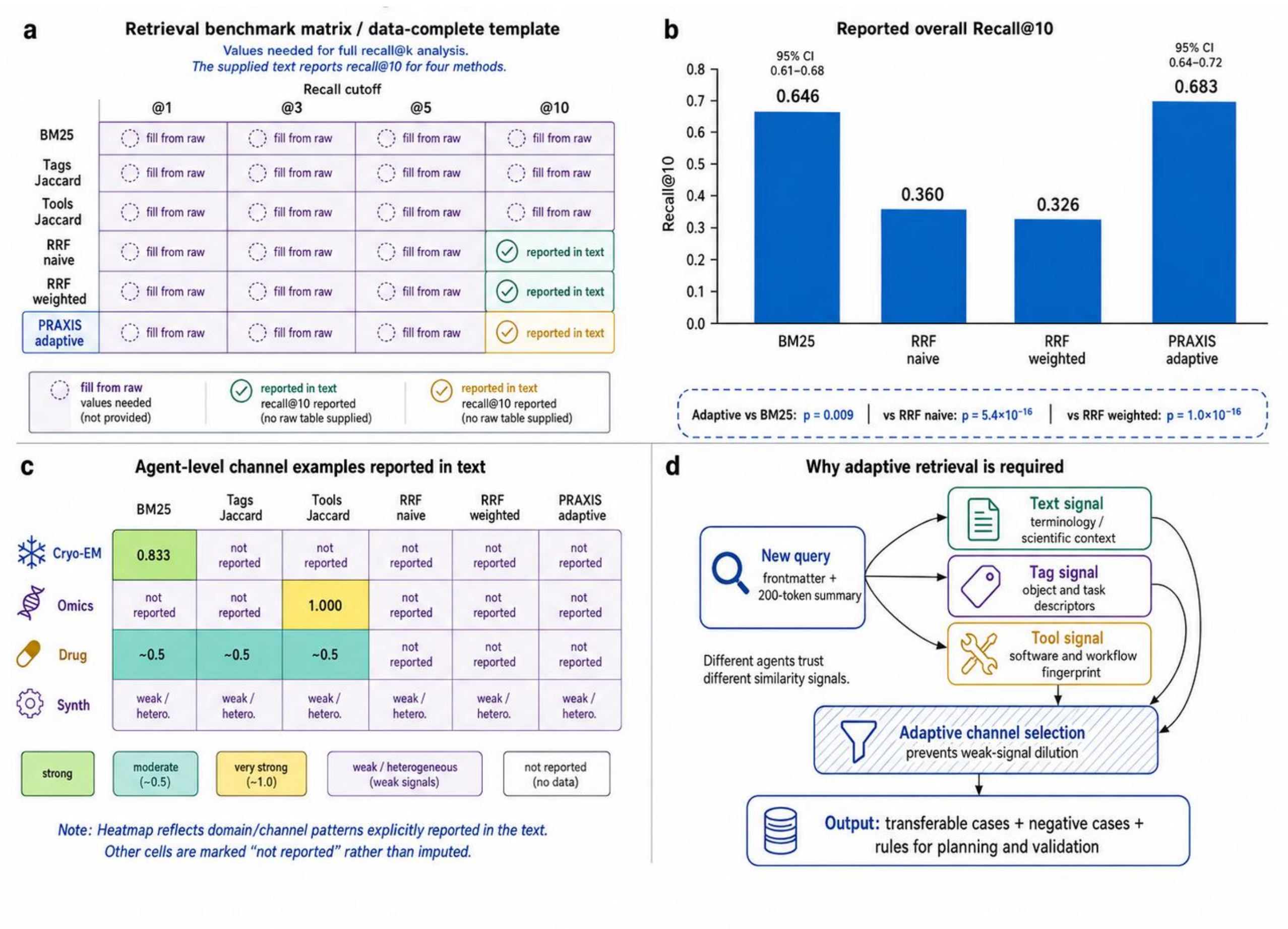

**Figure S1.** Complete results of the 110-case leave-one-out retrieval benchmark. The figure compares BM25, Tags Jaccard, Tools Jaccard, RRF naive, RRF weighted, and PRAXIS adaptive retrieval across recall@1, recall@3, recall@5, and recall@10, with bootstrap 95% confidence intervals and paired Wilcoxon test results. The accompanying heatmap shows agent-specific recall@10, highlighting distinct dominant retrieval channels across domains.

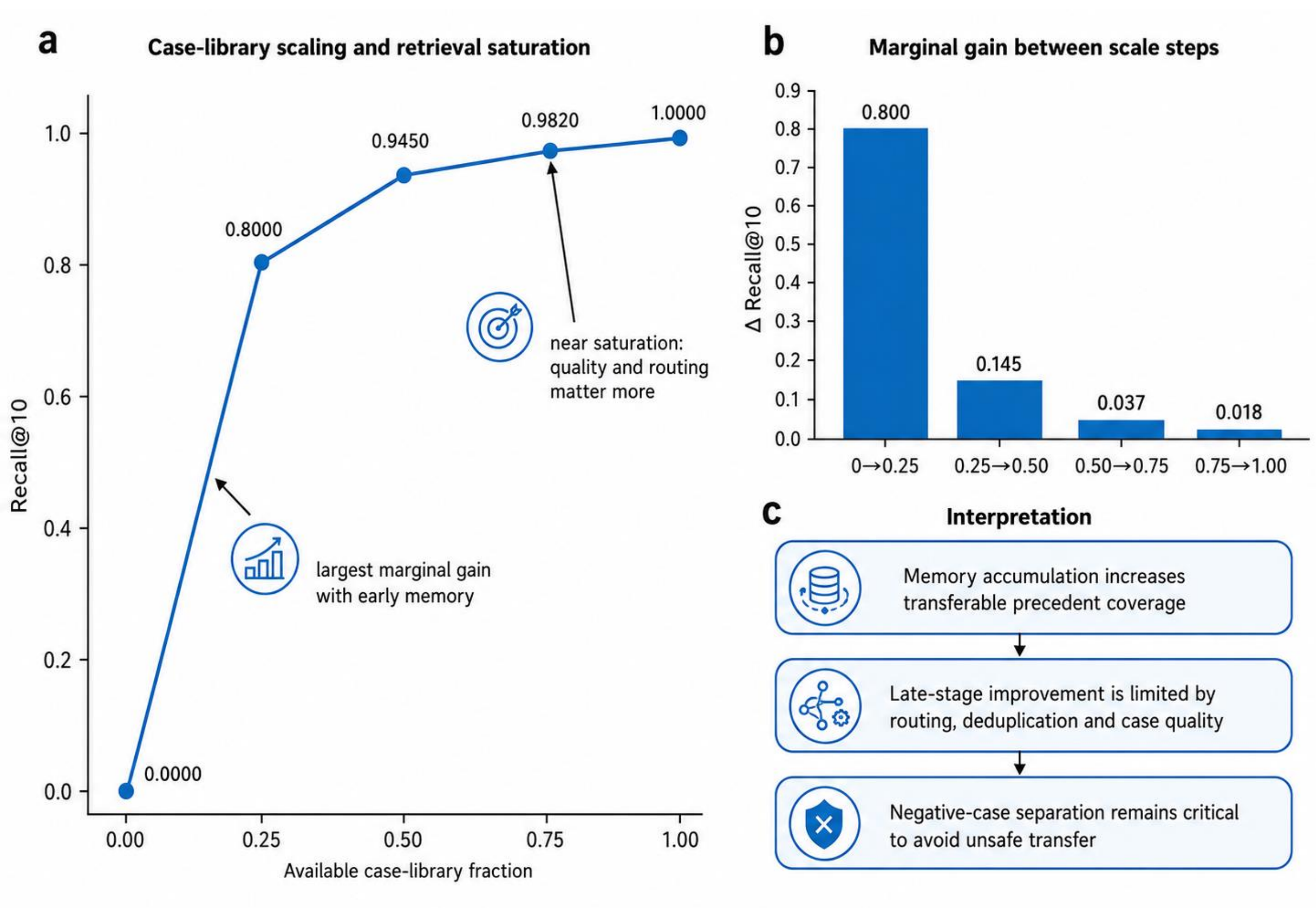

**Figure S2.** Case-library scaling and retrieval saturation. The figure shows recall@10 as the available case-library fraction increases from 0 to 0.25, 0.50, 0.75, and 1.00, with confidence intervals from repeated subsampling where applicable. Recall@10 increases monotonically with case-library size, but with diminishing marginal gains. This indicates that long-term case accumulation improves retrieval coverage, whereas further gains at larger scales depend more on retrieval routing, deduplication, negative-case separation, and case-quality control.

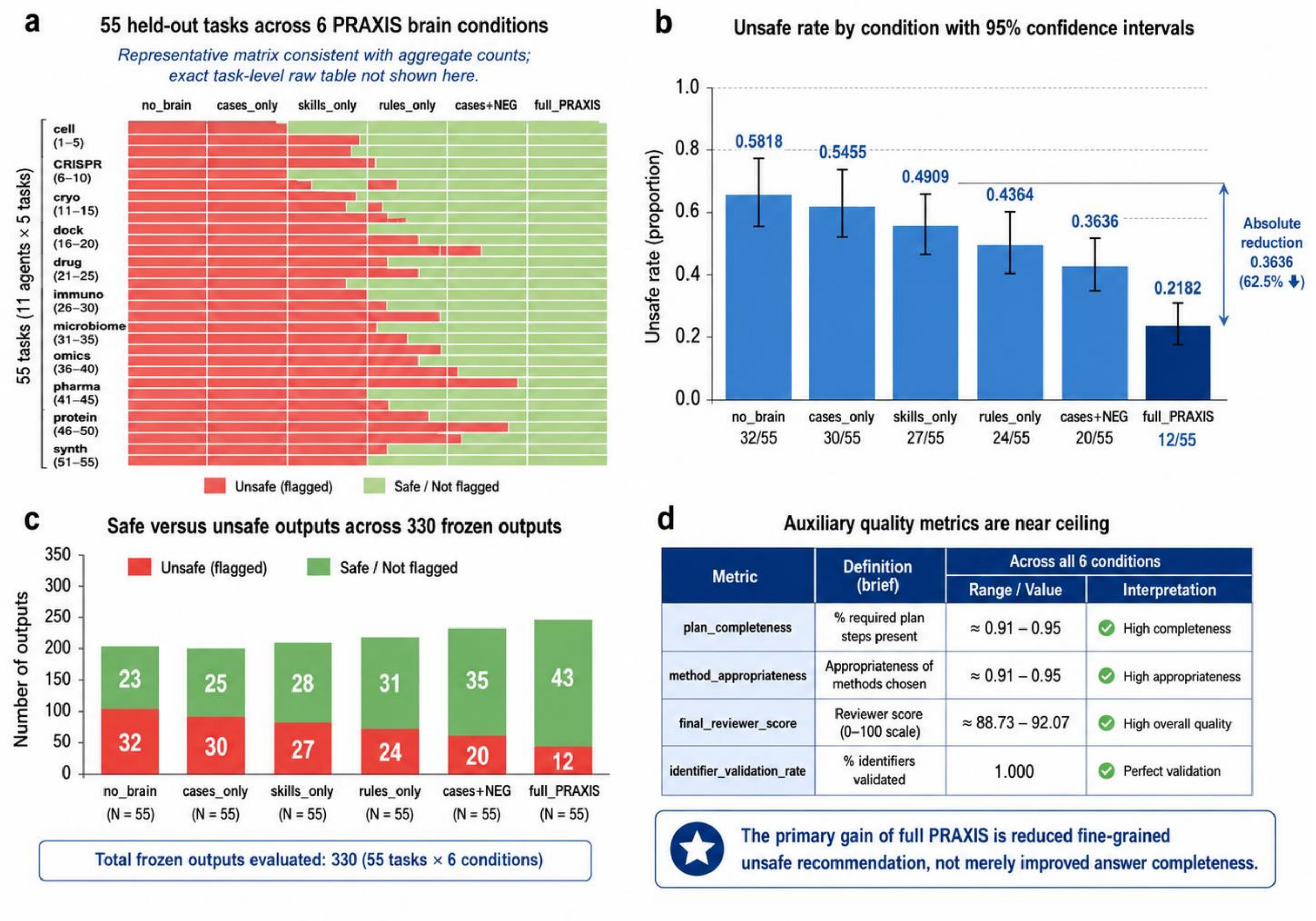

**Figure S3.** Complete unsafe-recommendation matrix across 55 held-out tasks and six PRAXIS brain conditions. This heatmap shows unsafe-recommendation calls for 55 tasks under no_brain, cases_only, skills_only, rules_only, cases+NEG, and full_PRAXIS. It also reports agent-level task groups, condition-specific unsafe_rate, bootstrap confidence intervals, and auxiliary metrics including plan_completeness, method_appropriateness, final_reviewer_score, and identifier_validation_rate.

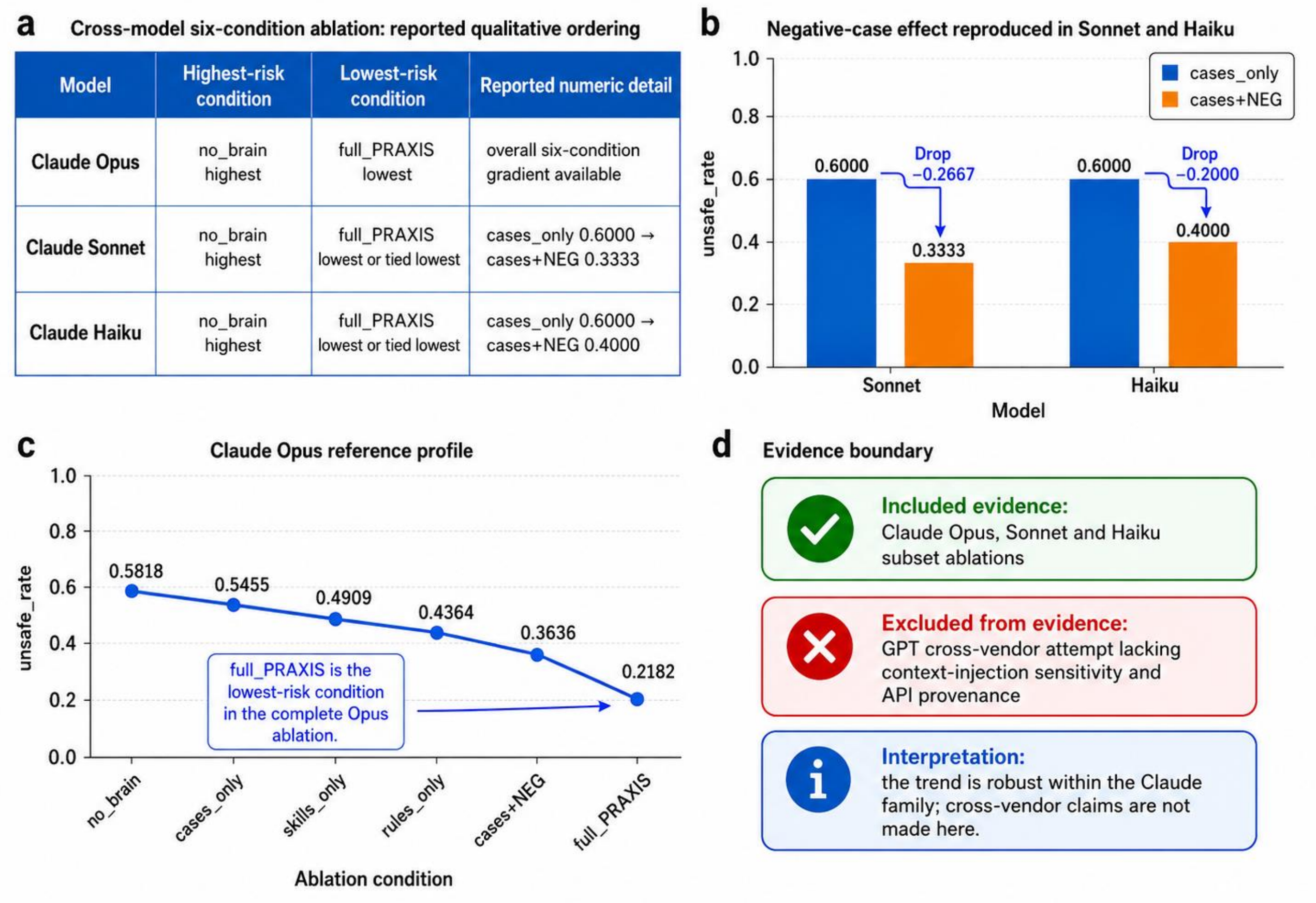

**Figure S4.** Cross-model ablation results on Claude Opus, Sonnet, and Haiku. This figure shows the six-condition ablation results for the cryo, drug, and omics agent subsets across Claude Opus, Claude Sonnet, and Claude Haiku. Each model was evaluated under no_brain, cases_only, skills_only, rules_only, cases+NEG, and full_PRAXIS conditions using unsafe_rate as the metric. Across all three models, no_brain showed the highest risk, whereas full_PRAXIS achieved the lowest or jointly lowest unsafe_rate. The reduction from cases_only to cases+NEG was also reproduced in Sonnet and Haiku, supporting the risk-suppression role of negative-case memory.